\documentclass[aps,ams,showpacs,superscriptaddress]{revtex4}

\usepackage{color}
\usepackage{graphicx,color}
\usepackage{latexsym}
\usepackage{amsmath,amssymb}        
\usepackage[draft=false]{hyperref}
\usepackage{natbib}
\usepackage{url}
\usepackage{psfrag}

\usepackage{mathrsfs}
\DeclareSymbolFontAlphabet{\mathrsfs}{rsfs}
\DeclareMathAlphabet{\mathcal}{OMS}{cmsy}{m}{n}

\newcommand{\scri}{\mathrsfs{I}}

\begin{document}


\title{Numerical solution of the wave equation on particular space-times using CMC slices and scri-fixing conformal compactification}


\author{A. Cruz-Osorio}
\affiliation{Instituto de F\'{\i}sica y Matem\'{a}ticas, Universidad
              Michoacana de San Nicol\'as de Hidalgo. Edificio C-3, Cd.
              Universitaria, 58040 Morelia, Michoac\'{a}n,
              M\'{e}xico.}

\author{A. Gonz\'alez-Ju\'arez}
\affiliation{Facultad de Ciencias F\'isico Matem\'aticas, Universidad Aut\'onoma de Puebla, 
Apartado Postal 1152 Puebla, 72001, Pue., Mexico. 
}

\author{F. S. Guzm\'an}
\affiliation{Instituto de F\'{\i}sica y Matem\'{a}ticas, Universidad
              Michoacana de San Nicol\'as de Hidalgo. Edificio C-3, Cd.
              Universitaria, 58040 Morelia, Michoac\'{a}n,
              M\'{e}xico.}

\author{F. D. Lora-Clavijo}
\affiliation{Instituto de F\'{\i}sica y Matem\'{a}ticas, Universidad
              Michoacana de San Nicol\'as de Hidalgo. Edificio C-3, Cd.
              Universitaria, 58040 Morelia, Michoac\'{a}n,
              M\'{e}xico.}


\date{\today}


\begin{abstract}
In this paper we present in detail the numerical solution of the conformally invariant wave equation on top of a fixed background space-time corresponding to two different cases: 
i) 1+1 Minkowski space-time in Cartesian coordinates and ii) Schwarzschild space-time. In both cases we use hyperboloidal constant mean curvature slices and scri-fixing conformal compactification, and solve the wave equation on the conformal space-time. In the case of the Schwarzschild space-time we study the quasinormal mode oscillations and the late-time polynomial tail decay exponents corresponding to a mass-less scalar field. We also present general formulas to construct hyperboloidal constant mean curvature slicings of spherically symmetric, static, space-times in spherical coordinates.
\end{abstract}




\maketitle



\section{Introduction}

A common problem in numerical relativity is the global treatment of asymptotically flat space-times, specially related to the implementation of boundary conditions on an artificial time-like boundary and the measurement of gravitational radiation by detectors located on time-like trajectories. 
Most of the initial value problems formulated are based on a Cauchy space-like foliation with the 
hypersurfaces approaching spatial infinity $i^0$. Then, boundary conditions must be supplied at an artificial 
boundary, for instance, absorbing boundary conditions \cite{sarbach}. Recent efforts point to the use of hyperboloidal foliations of the space-time, because asymptotically the spatial slices reach future null infinity $\scri^{+}$
and for asymptotically flat space-times, the gravitational radiation is well defined at such boundary \cite{bondi,sachs}, so that the measurement of the gravitational radiation can be observed by detectors near or at that boundary. More common efforts are related to matching techniques of Cauchy slices with slices approaching future null infinity \cite{stewart,Frie,Frau}, which has already been applied to the binary black hole problem \cite{Pollney2009}. On the other hand, it is desirable to compactify the space-time foliated in this manner in order to work on a numerical domain that contains $\scri^{+}$, for which it is necessary to compactify the spatial coordinates in an appropriate way such that the space-time is regular there, at future null infinity. 
At present time, foliations with boundaries at future null infinity are being used for various applications, like the solution of perturbation equations \cite{whs,dario,anil2009} and the study of tails \cite{tiglio}.

The idea behind scri-fixing conformal compactification is the following: i) use hyperboloidal slices that reach $\scri^{+}$, ii) compactify the spatial coordinate, iii) rescale the metric with an appropriate conformal factor. We proceed by defining a new time coordinate $t = \tilde{t} - h(\tilde{y})$, where ($\tilde{t},\tilde{y}$) are the original time and space coordinates 
This transformation has the advantage that keeps the time direction invariant. That is, regardless of the choice of the $h(\tilde{y})$,  the time-like killing vector has the same form in both coordinate systems. A compactifying coordinate is introduced in the form  $\tilde{y} =  \frac{y}{\Omega}$, where $\Omega=\Omega(y)$ is a non-negative conformal factor that vanishes at  $\scri^{+}$ \cite{anil1}. Then the physical metric $\tilde{g}$ becomes singular at the boundary of the compact coordinate. In order to remove the singularities  the metric is rescaled $g=\Omega^{2}\tilde{g}$ with the conformal factor so that the conformal metric $g$ is regular at the boundary $\scri^{+}$. On the other hand, for the construction of $h(\tilde{y})$, we restrict in this paper to the case of constant mean curvature (CMC) slices \cite{Murc}. 

In this manuscript we present the solution of the wave equation for the extended Minkowski and Schwarzschild space-times described with CMC slices, construct the foliations and compactification, and solve the wave equation on top of the resulting space-times with a twofold goal, on the one hand we verify as an academic exercise that the behavior of the wave function is consistent with the properties of the particular space-times constructed here, we are explicit in the construction of the numerical solution of the wave equation in one spatial dimension and illustrate the role played by the gauge. On the other hand, we study in particular the wave function on top of the Schwarzschild space-time, and track the quasinormal mode frequencies and tail decay rates of the amplitude of the wave function, which is also the case of a test mass-less scalar field evolving on the space-time.

Once the space-time has been foliated with hyperboloidal slices, the spatial coordinate has been compactified and the metric rescaled, it is necessary to solve a wave equation that is physically meaningful. Thus the wave equation to be solved has to be a conformally invariant one. Considering the conformally rescaled metric $g=\Omega^{2} \tilde{g}$, the following identity applies for an $n$-dimensional space-time \cite{wald}

\begin{equation}
\left[ \Box -\frac{n-2}{4(n-1)} R \right] \phi = \Omega^{-3} \left[ \tilde \Box  - \frac{n-2}{4(n-1)} \tilde{R} \right] \tilde{\phi} =0,\label{eq:nwave}
\end{equation}

\noindent which means that the operator in brackets is invariant under conformal transformations and is the one to be considered as the conformally invariant D'Alambertian operator. In this expression $R$ and $\tilde R$ are the Ricci scalar in the conformal and physical space-times respectively,  $\phi= \frac{\tilde{\phi}}{\Omega}$ is the conformal wave function, $\tilde{\phi}$ the physical wave function and $\Box$ is defined by

\begin{equation}
 \Box \phi = \frac{1}{\sqrt{-g}}\partial_{\mu} [\sqrt{-g}g^{\mu\nu}\partial_{\nu}\phi],\label{eq:invwave}
\end{equation}

\noindent where $g_{\mu\nu}$ is the conformal metric and $g$ its determinant. For the cases presented in this manuscript, the conformally invariant wave equation for the 1+1 Minkowski space-time is
$\Box \phi =0$, whereas for the Schwarschild space-time ($n=4$) the conformally invariant wave equation reads $\Box\phi -\frac{1}{6} R \phi =0$.

We solve these two equations as initial value problems in each case using a first order variable formulation. In the case of the solution of the wave equation on the Schwarzschild space-time, the equation also corresponds to a test mass-less scalar field, which is related to quasinormal mode oscillations of the black hole and study the polynomial tail decay of the amplitude of the scalar field; in such case we use mass-less scalar field and wave function indistinctly.

This paper is organized as follows. In section II we solve the conformally invariant wave equation on top of the 1+1 Minkowski space-time, and provide diagrams illustrating the results. In section III we solve the conformally invariant wave equation on top of a scri-fixing conformally compactified version of the Schwarzschild solution; we show the quasinormal mode oscillations and the tail decay of the scalar field, and compare with previous results. In section IV we present general formulas useful to set up slicings with constant mean curvature and scri-fixing conformal compactification for spherically symmetric and static space-times in spherical coordinates; we show as an example the 3+1 Minkowski space-time in spherical coordinates. Finally in section V we summarize the main results.


\section{ 1+1 Minkowski space-time}

\subsection{ Foliation and scri-fixing}
\label{sec:foliation}
We choose the coordinates of the space-time $x^{\mu}=(\tilde{t}, \tilde{x})$ and the line element 

\begin{equation}
 d\tilde{s}^2 = -d\tilde{t}^{2} + d\tilde{x}^{2}, \label{eq:m1p1}
\end{equation}

\noindent where $\tilde{t},\tilde{x}\in(-\infty,+\infty)\times (-\infty,+\infty)$.
In order to construct the hyperboloidal slices, we perform a time transformation
$t = \tilde{t}-h(\tilde{x})$. Using this new time coordinate, the metric reduces to

\begin{equation}
d\tilde{s}^2 = -dt^{2} - 2h'dtd\tilde{x} + [1-h^{'2}]d\tilde{x}^{2}.
\end{equation}

\noindent where $h^{\prime}=\frac{dh}{d\tilde{x}}$. In order to have a simple interpretation of the injection of hypersurfaces into the space-time we compare this expression with the usual  ADM-like metric 

\begin{equation}
d\tilde{s}^2 = \left( -\tilde{\alpha}^2  + \tilde{\gamma}^{2} \tilde{\beta}^2 
\right)  dt^2 + 2  \tilde{\beta}\tilde{\gamma}^2dt 
d\tilde{x}+ \tilde{\gamma}^2 d\tilde{x}^2, \label{eq:metricadm1p1} 
\end{equation}

\noindent and we read off the lapse function $\tilde \alpha = \frac{1}{\sqrt{1-h^{'2}}}$, and then the future-pointing unit vector normal to the hypersurface $n^{\mu}$ is

\begin{eqnarray}
n_{\mu}&=& (-\tilde\alpha,0) = \left(-\frac{1}{\sqrt{1-h^{'2}}},0\right), \nonumber\\
n^{\mu}&=&\left(\sqrt{1-h^{'2}}, \frac{h'}{\sqrt{1-h^{'2}}} \right), \label{eq:norvec}
\end{eqnarray}

\noindent with which it is possible to calculate the mean extrinsic curvature of the slice $\tilde{k} = \nabla_{\mu}n^{\mu} = \frac{1}{\sqrt{-g}}\partial_{\mu}(\sqrt{-g}n^{\mu})$:

\begin{equation}
\tilde{k}=\partial_{\tilde{x}}\left( \frac{h'}{\sqrt{1-h^{'2}}} \right),~~~~ \Rightarrow
\tilde{k}\tilde{x}=\left( \frac{h'}{\sqrt{1-h^{'2}}} \right) + C,
\end{equation}

\noindent where the second equality is obtained after assuming $\tilde k$ is constant, and finally $C$ is an integration constant. This equation can be manipulated to give a complete description of the slices:

\begin{equation}
h^{\prime}= \frac{\tilde{k}\tilde{x}}{\sqrt{1+(\tilde{k}\tilde{x})^{2}}} ~~\Rightarrow ~~
h(\tilde{x})= \sqrt{a^{2}+ \tilde{x}^{2}},
\end{equation}

\noindent where $a=\frac{1}{\tilde{k}}$. The space-time in these new coordinates is described with hyperboloidal constant mean curvature slices.

In order to compactify the space-time at the future null infinity boundary it suffices to define a new spatial compact coordinate $x$ defined through $\tilde{x}=\frac{x}{\Omega}$. A convenient choice of the conformal factor is $\Omega=1-x^2$, because we want the two asymptotic ends $\tilde x \rightarrow \pm \infty$ correspond to $x=\pm 1$ in the compactified coordinate. This choice has been used for instance in \cite{whs} to construct the solution of the perturbation equation of charged wormholes. In this way, using $\tilde{x}=x/\Omega$ with $\Omega=1-x^2$, 
$h(\tilde{x})=\sqrt{a^2+\tilde{x}^2}$, implies $h^{\prime} = \frac{x}{\sqrt{a^2 (1-x^2)^2+x^2}}$, and finally  
the conformal metric reads

\begin{eqnarray}
ds^2 &=&  -(1-x^2)^2dt^2 - \frac{2x(1+x^2)}{\sqrt{a^2 (1-x^2)^2 + x^2}}dtdx \nonumber\\
	&+& \frac{a^2 (1+x^2)^2}{a^2 (1-x^2)^2 + x^2}dx^2.
	\label{eq:metric_mink_conformal}
\end{eqnarray}

\noindent In order to find the gauge functions, this conformal metric has to be identified with the standard 1+1 metric of type (\ref{eq:metricadm1p1}), where we remove the tildes because we deal with the conformal metric. The identification implies the following gauge functions:

\begin{eqnarray}
\alpha^2 &=& \frac{x^2}{a^2} + (1-x^2)^2,\nonumber\\
\beta   &=& -\frac{x}{a^2}\frac{\sqrt{a^2 (1-x^2) + x^2}}{1+x^2},\nonumber\\
\gamma^2 &=& \frac{a^2 (1+x^2)^2}{a^2 (1-x^2)^2 + x^2},\label{eq:gauge_Mink_wave}
\end{eqnarray}

\noindent which are the ones used to describe the space-time on top of which we solve the wave equation.


\subsection{The wave equation}
\label{sec:wave}

We solve the conformally invariant wave equation. For this case $n=2$ and $R=0$, the wave operator is conformally invariant, thus we solve the equation (\ref{eq:invwave})

\begin{equation}
 \Box \phi = \frac{1}{\sqrt{-g}}\partial_{\mu} [\sqrt{-g}g^{\mu\nu}\partial_{\nu}\phi]=0 \nonumber
\end{equation}
\noindent
for the metric (\ref{eq:metric_mink_conformal}, \ref{eq:gauge_Mink_wave}) as an initial value problem.

It is straightforward to show that assuming $\alpha=\alpha(x)$, $\beta=\beta(x)$ and $\gamma=\gamma(x)$, the wave equation can be written as a first order system of equations given by

\begin{eqnarray}
\partial_t \psi&=&  \partial_x\left( \frac{\alpha}{\gamma}\pi + \beta\psi \right),\nonumber\\
\partial_t \pi&=&  \partial_x \left(\beta \pi + \frac{\alpha}{\gamma}\psi \right), \nonumber\\
\partial_x \phi &=& \psi,
\label{eq:we1st}
\end{eqnarray}

\noindent where $\pi=\frac{\gamma}{\alpha}\partial_t \phi - \frac{\gamma}{\alpha}\beta \partial_x \phi$ and $\psi=\partial_x \phi$. This system of equations holds as long as $\alpha \gamma \ne 0$. The third equation is a definition of $\psi$ but it is also a constraint of the system that has to be satisfied. The value of the wave function is obtained from the definition of $\pi$, that is $\partial_t \phi= \frac{\alpha}{\gamma}\pi + \beta\psi$ and can be integrated in time at the same time as $\pi$ and $\psi$.
We define the state vector ${\bf u} = (\pi,\psi)^T$, so that the wave equation can be cast in a balance law form $\partial_t {\bf u} + {\bf A} \partial_x {\bf u} = - \partial_x ({\bf A}) {\bf u}$, where

\begin{equation}
 {\bf A} = -
        \left(
                \begin{array}{cc}
                \beta   & \alpha/\gamma \\
                \alpha/\gamma  & \beta
               \end{array}
        \right)
\end{equation}

\noindent and the characteristic speeds of the system are given by 

\begin{equation}
\frac{dx}{dt}=\lambda_{\pm}=-\beta \pm \frac{\alpha}{\gamma}.\label{eq:speeds}
\end{equation}

\noindent These eigenvalues are real and distinct and in fact correspond to a complete set of eigenvectors. Then, this system of equations is symmetric hyperbolic, and together with initial data for $\psi$ and $\pi$ it corresponds to a well posed initial value problem. We define such problem in the domain $x \in [-1,1]$, $t \in [0,\infty)$.  We choose initial data corresponding to an initially time-symmetric wave function with Gaussian profile. In particular we use $\phi = e^{-x^2/(0.1)^2}$, $\psi = \partial_x \phi$ and $\pi=0$.

We solve this problem using a finite differences approximation, with a method of lines using second order stencils along the spatial direction and a third order Runge-Kutta time integrator. 
The question about the boundary conditions $\pi(-1,t)$, $\psi(-1,t)$, $\pi(1,t)$ and $\pi(1,t)$ remains, and the answer is that there is no need to apply boundary conditions (e.g. radiative boundary conditions) because we notice that in the continuum limit at $x=-1$ the characteristic speed corresponding to the mode moving to the right is zero ($\lambda_{+}=0$), whereas the characteristic speed of the mode moving to the left at $x=1$ is also zero ($\lambda_{-}=0$), and then no signals are propagated into the domain of the problem.

\begin{figure*}[ht]
\includegraphics[width=5.5cm]{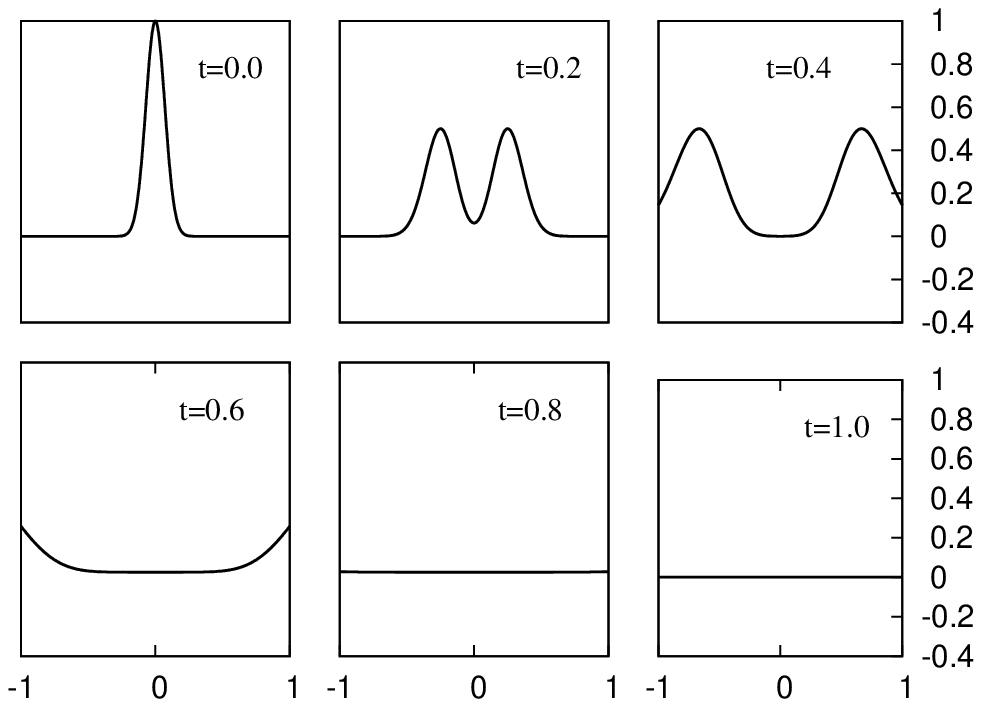}
\includegraphics[width=5.5cm]{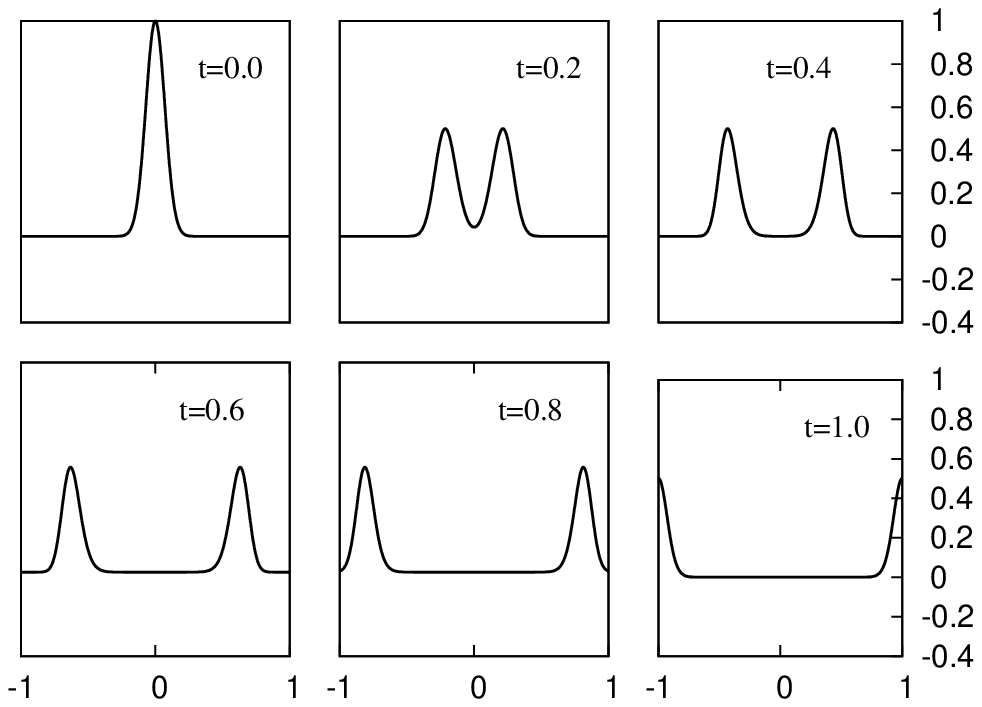}
\includegraphics[width=5.5cm]{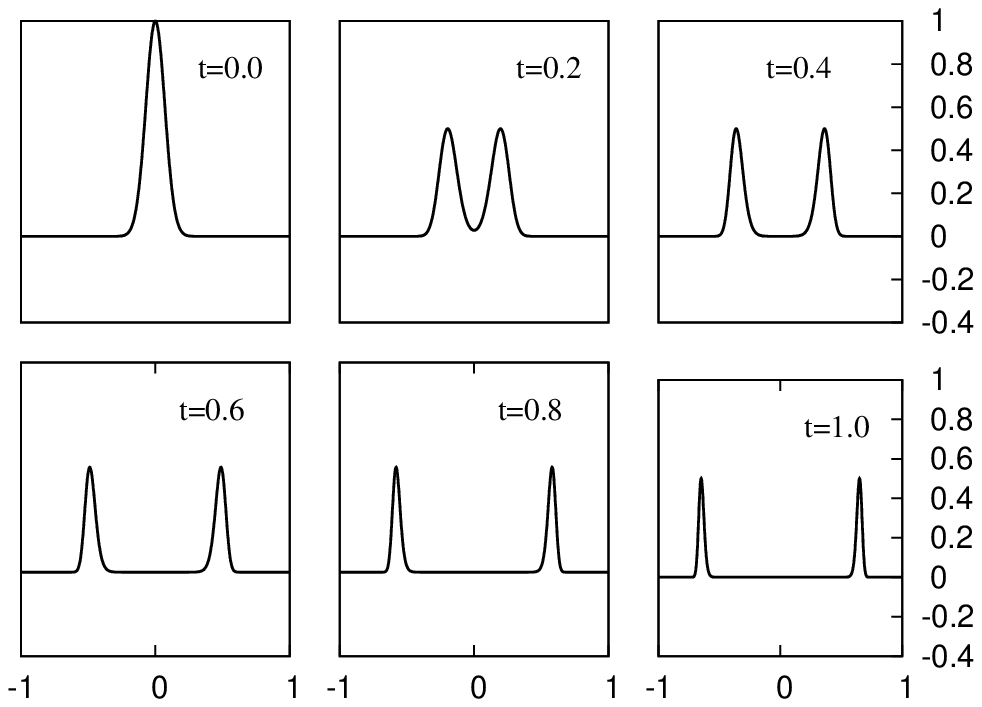}
\caption{\label{fig:snaps_Mink1p1} Evolution of the wave function for $a=0.5$ (left), $a=1$ (middle) and $a=5$ (right) or equivalently $\tilde{k}=2,1,0.2$ respectively. These results are consistent with the space-time diagrams in Fig. \ref{fig:brutal_Mink}, where  the light cones seem to get narrow near the boundaries for big values of $a$ (small values of the curvature) which explains why the pulses when they approach the boundaries slow down and shrink.}
\end{figure*}

\begin{figure*}[ht]
\includegraphics[width=7.5cm]{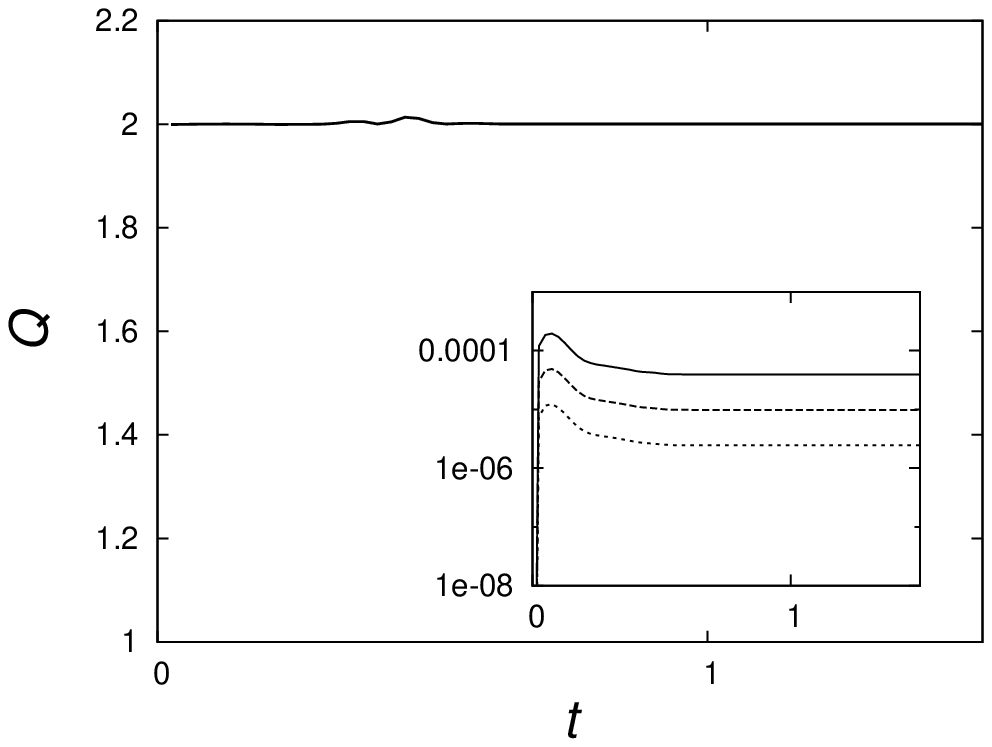}
\includegraphics[width=7.5cm]{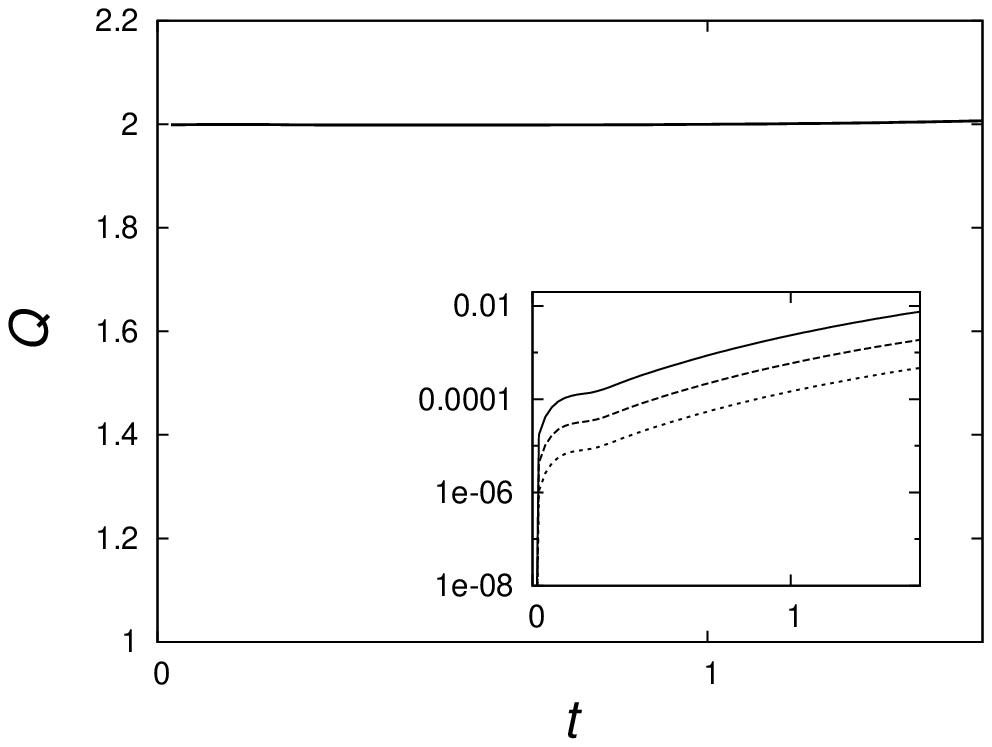}
\includegraphics[width=7.5cm]{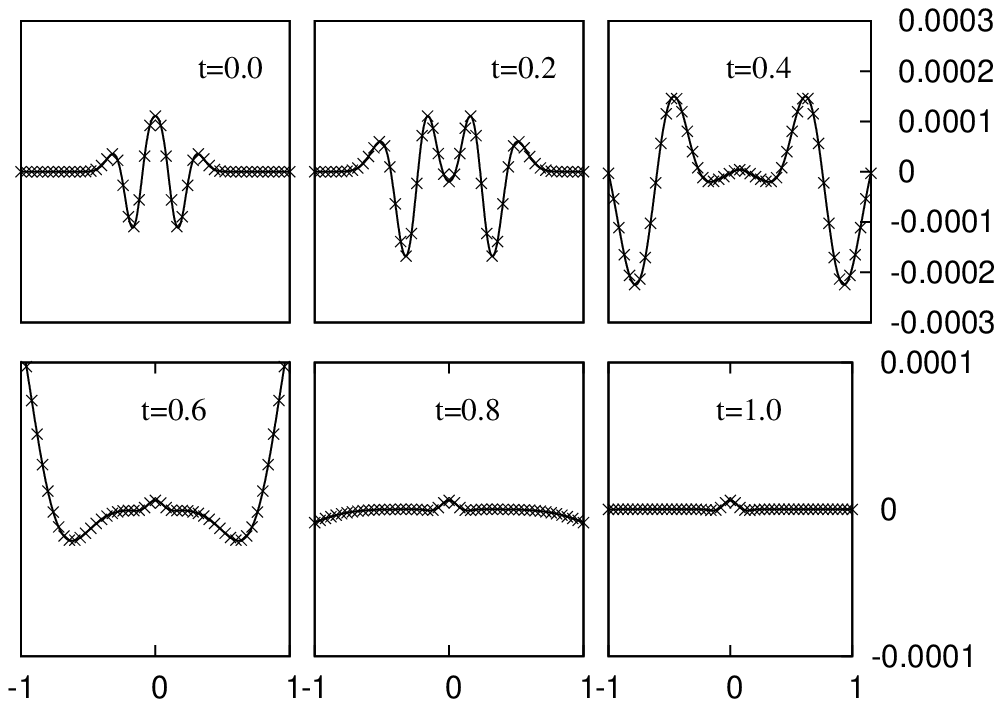}
\includegraphics[width=7.5cm]{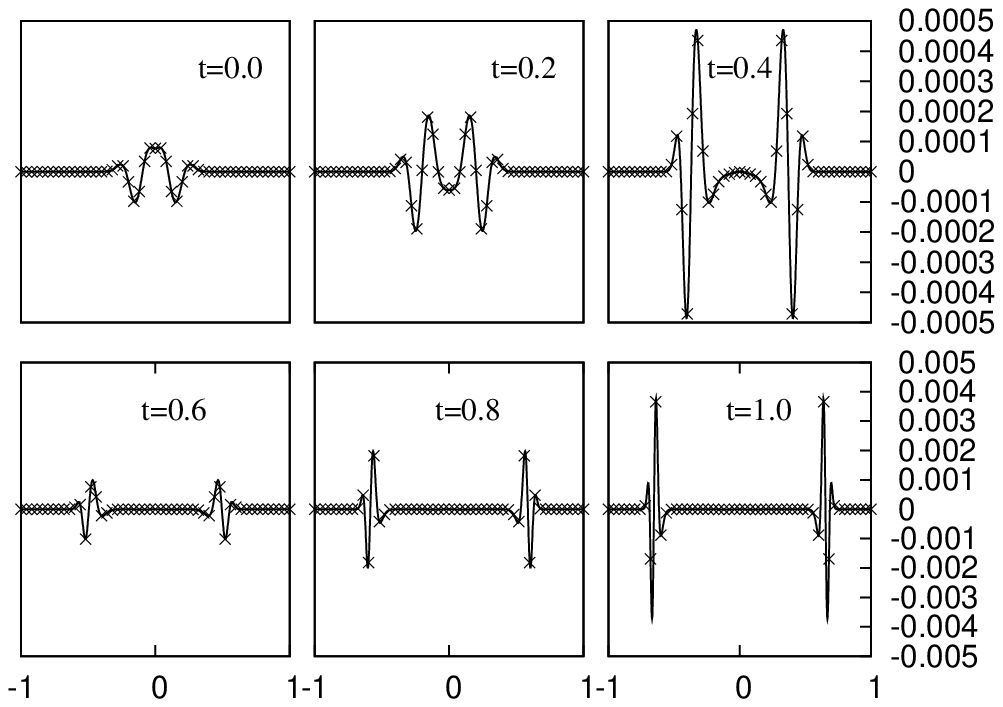}
\caption{\label{fig:convergencem1p1} Convergence tests for the solution of the wave equation with $\tilde{k}=2,0.2$ or equivalently $a=0.5,5$ on the left and right panels respectively. We use three resolutions $\Delta x=0.002,0.001,0.0005$. (Top) We show the order of convergence $Q$ of the constraint $C=\psi - \partial_x \phi$ to zero, defined by $L_2(C)[using~medium~resolution]/L_2(C)[using~ high~resolution] = 2^Q$ because the ratio between resolutions is 2. In the insets we show the $L_2$ norm of the constraint for the three resolutions. (Bottom) Continuous lines represent the difference of $\phi$ for the two first resolutions and  $ \times-$lines represent the difference of $\phi$ for the two last resolutions multiplied by 4; the fact that the curves lie on top of each other indicates the second order self-convergence of $\phi$ for the snapshots presented. In the case $a=5$ (small curvature) we can see that the errors are not as small as in the other case; this is due to the fact that the pulses are being squeezed when approaching the boundaries, and then the resolution used is not enough to resolve the pulses, and then the errors are bigger, however we maintain second the convergence of second order.}
\end{figure*}

The result is as expected, the initial Gaussian pulse will split into two pulses with half the initial amplitude and opposite directions, and the speed of propagation of each pulse depends on the coordinates chosen. The results of the evolution for three different values of the mean curvature $\tilde{k}$ are shown in Fig. \ref{fig:snaps_Mink1p1}, where the effects of the different gauge choices are illustrated. In Fig. \ref{fig:convergencem1p1} we show the self-convergence of the solution, and the convergence of the constraint $\partial_x \phi= \psi$, which validates our numerical results.

\subsection{Space-time and conformal diagrams}
\label{sec:conformal}

In order to better illustrate the structure of the space-time  we construct the light cone structure and conformal diagrams, for which we first estimate the radial null geodesics by solving the 
condition $ds=0$:

\begin{equation}
\frac{dt}{dx} = -\frac{x(1+x^2)}{(1-x^2)^2\sqrt{a^2 (1-x^2)^2+x^2}} \pm 
	\frac{1+x^2}{(1-x^2)^2},\nonumber
\end{equation}

\noindent whose solution is

\begin{equation}
t = t_0 + \frac{-\sqrt{a^2 (1-x^2)^2 + x^2} \pm x}{1-x^2}.
\label{eq:mink_second_omega}
\end{equation}

\begin{figure*}[ht]
\includegraphics[width=5cm]{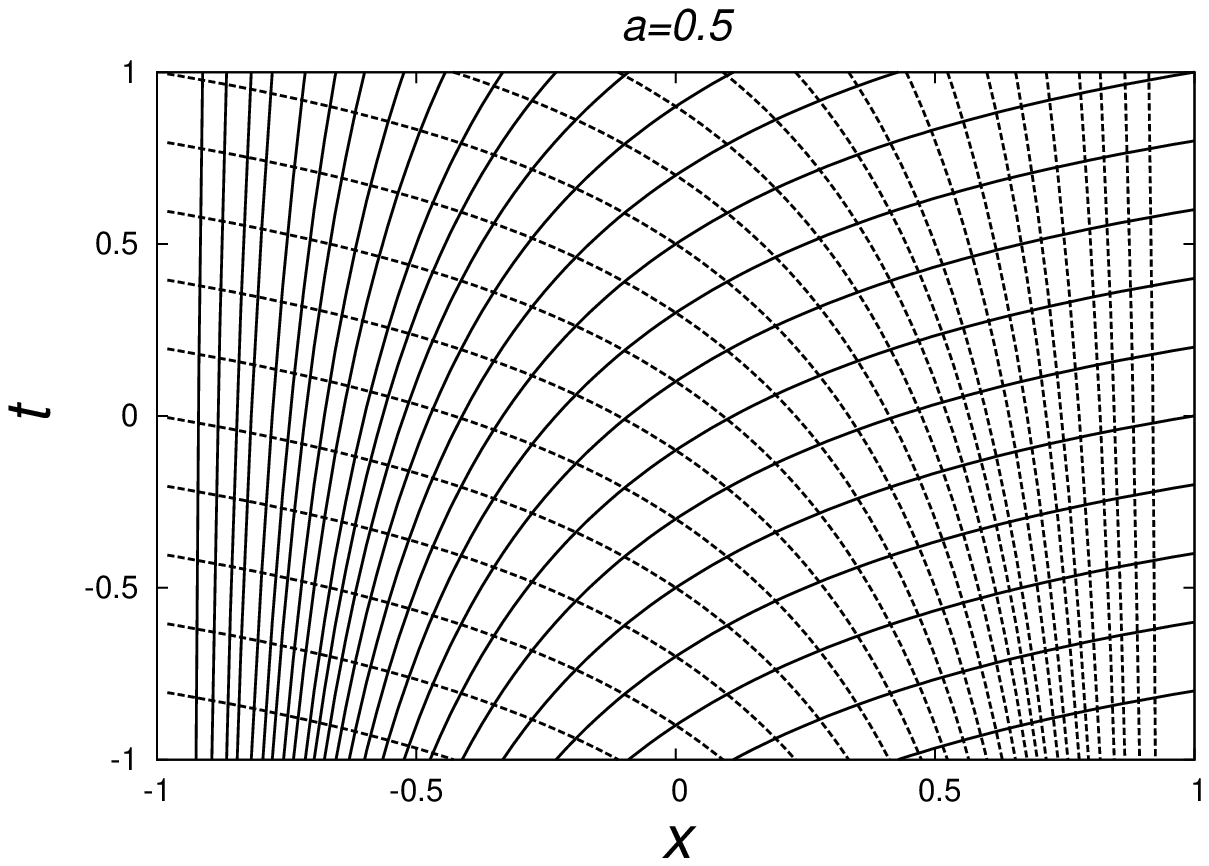}
\includegraphics[width=5cm]{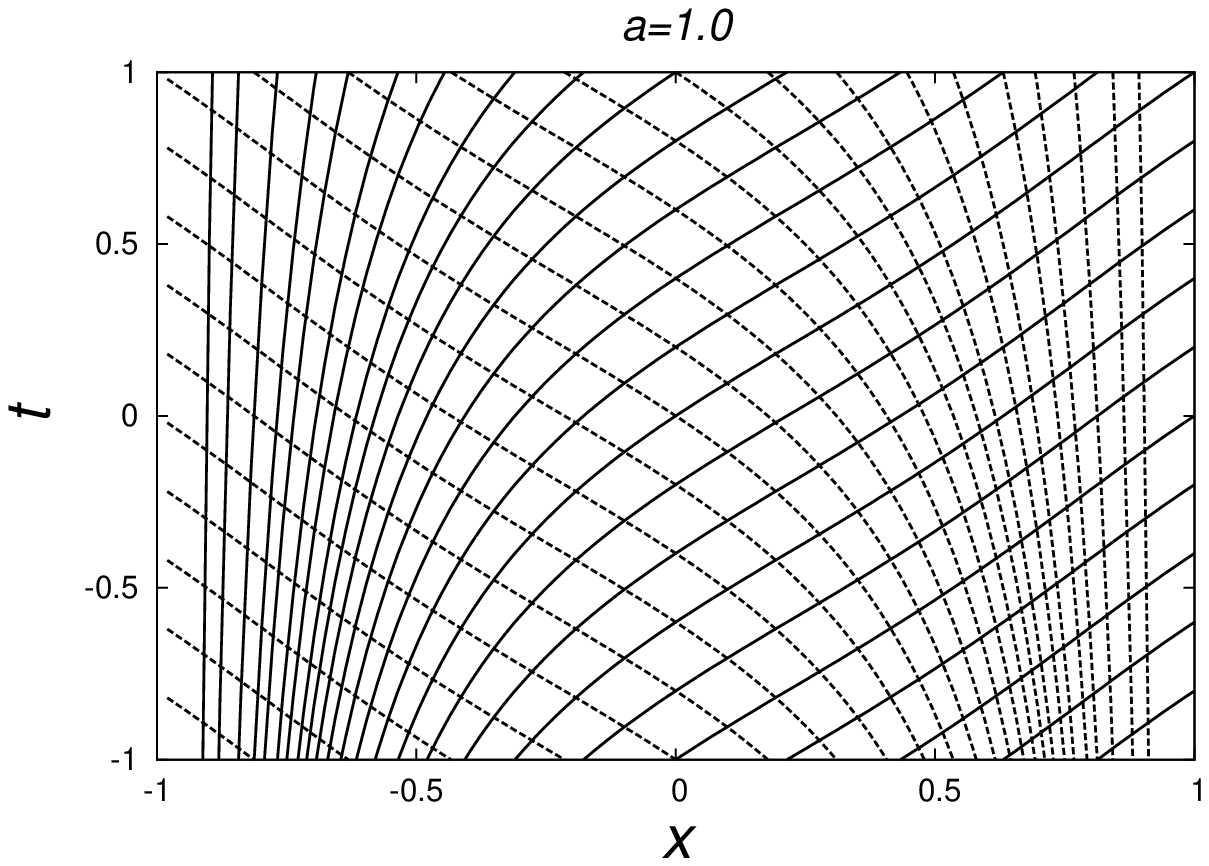}
\includegraphics[width=5cm]{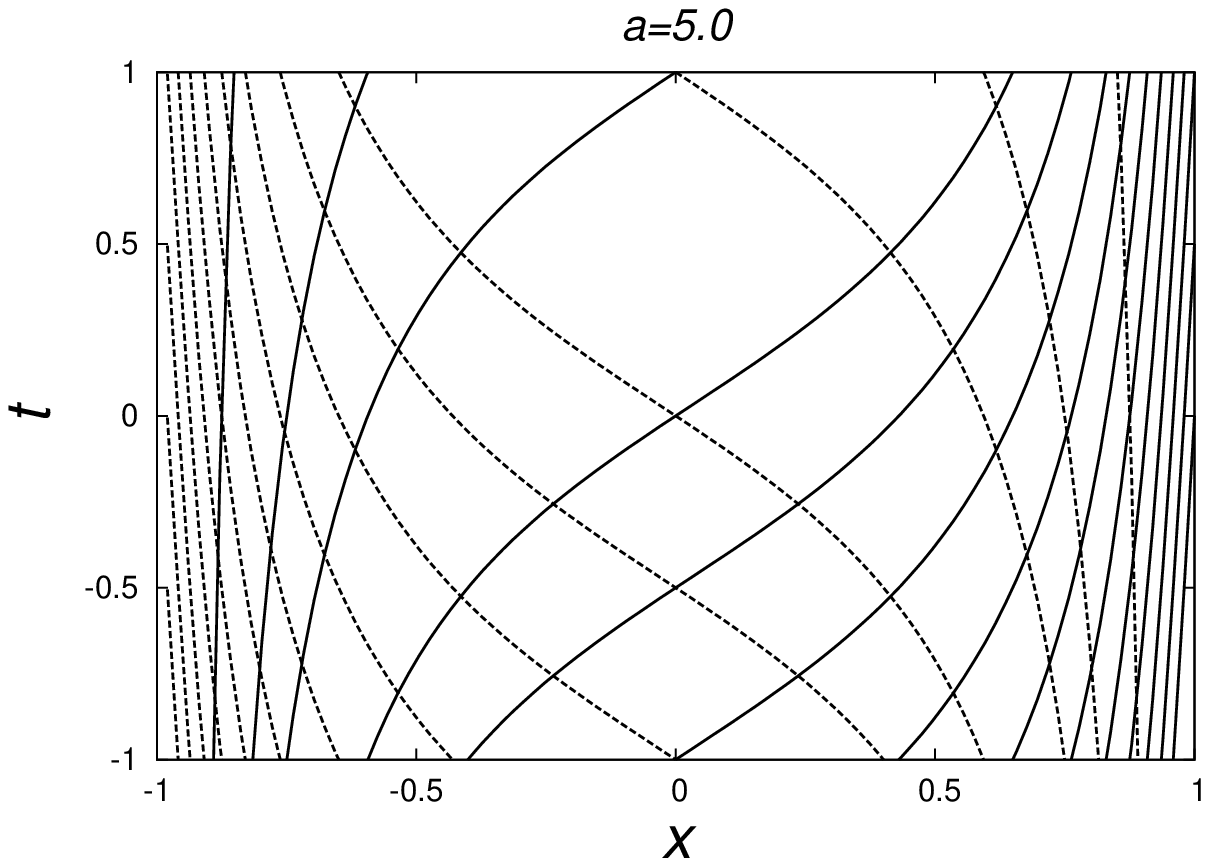}
\caption{\label{fig:brutal_Mink1p1} Space-time diagrams for slicings with $\tilde{k}=2,1,0.2$ or equivalently $a=0.5,1,5$. Continuous (dashed) lines represent null rays moving toward $x=+1$ ($x=-1$). These plots are constructed using (\ref{eq:mink_second_omega}) with various values of $t_0$.}
\end{figure*}

\noindent The result is shown in Fig. \ref{fig:brutal_Mink1p1}. What can be learned from this figure is that for small values of $\tilde{k}$ the light cones become narrow near $x=\pm 1$.  The implication is that the pulses of the initial data propagating outwards  will slow down and squeeze when they approach the boundaries in the case of small $\tilde{k}$.

\begin{figure*}[ht]
  \centering 
    \psfrag{ip}{$i^+$} \psfrag{im}{$i^-$} \psfrag{i0}{$i^0$}
     \psfrag{scrp}{$\scri^+$}\psfrag{scrm}{$\scri^-$}
\includegraphics[width=5cm]{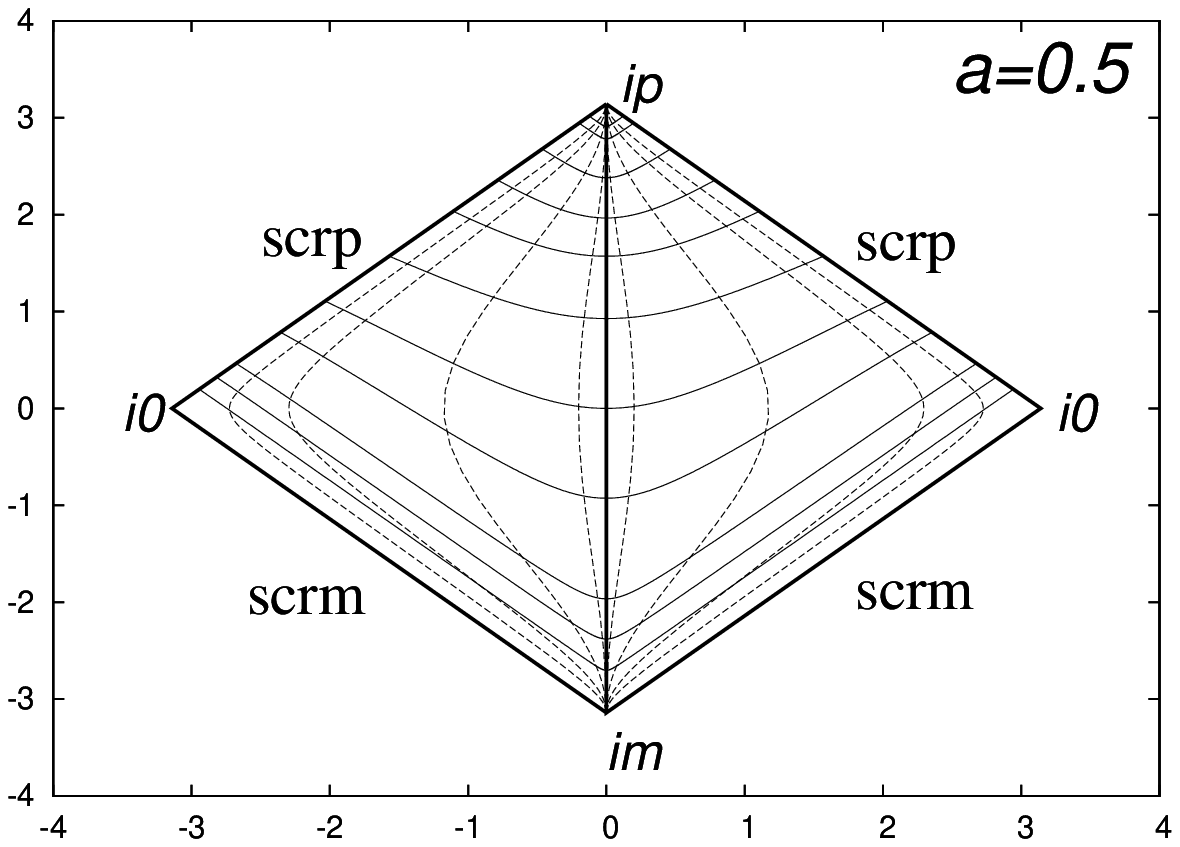}
\includegraphics[width=5cm]{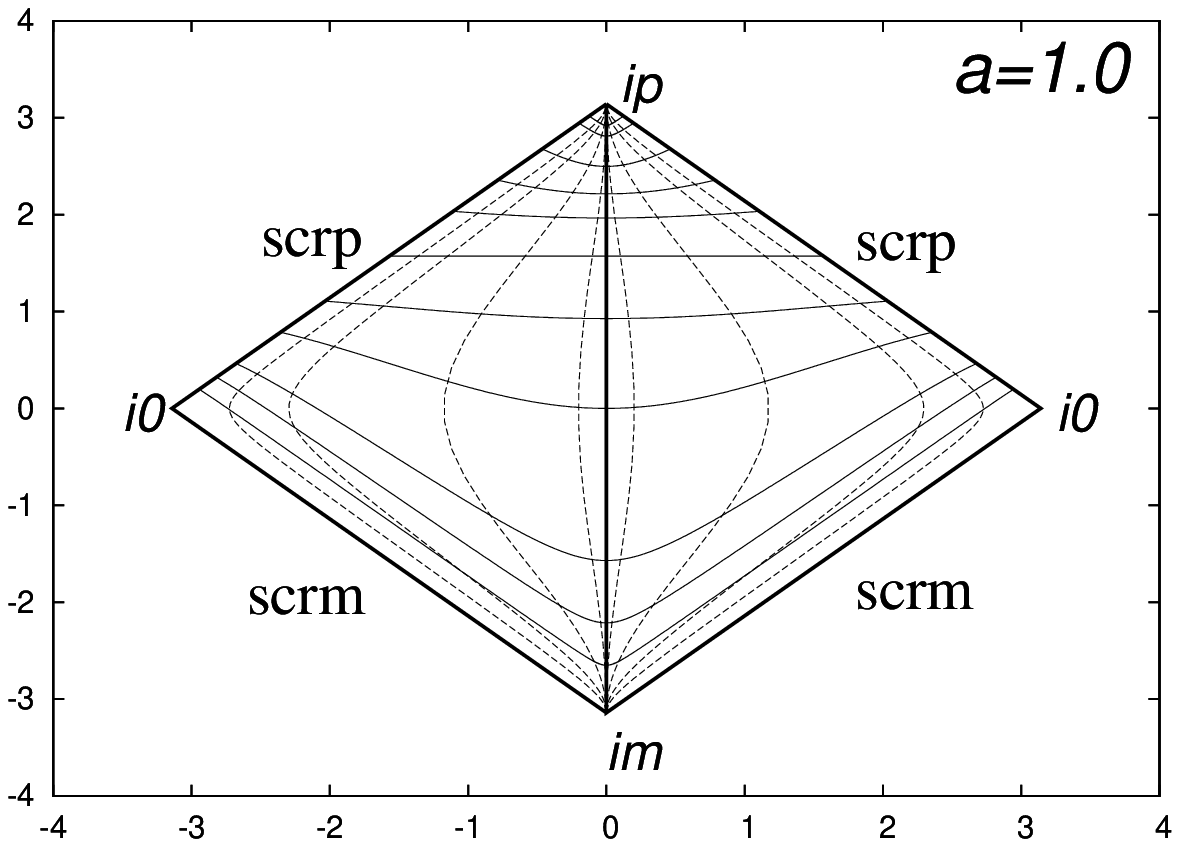}
\includegraphics[width=5cm]{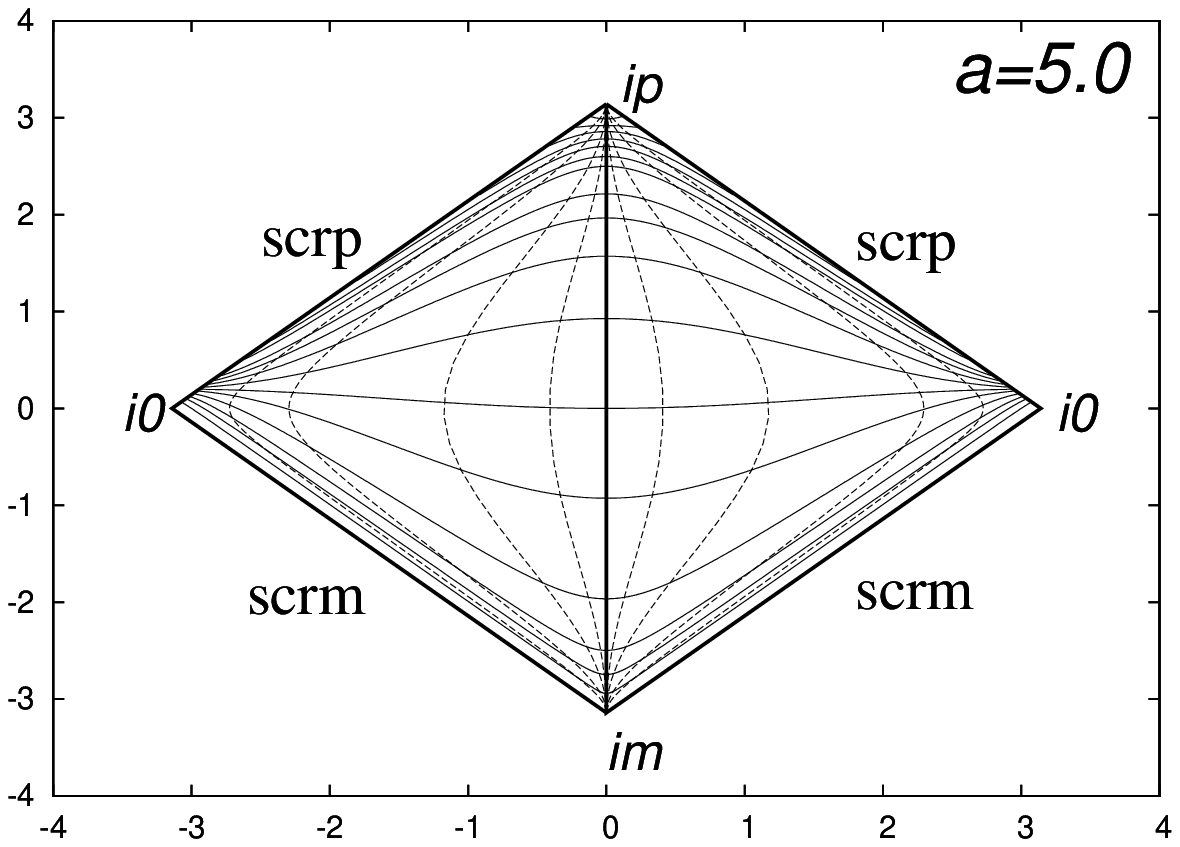}
\caption{\label{fig:conformal_Mink1p1} Conformal diagrams with hyperboloidal slices for $\tilde{k}=2,1,0.2$ or equivalently $a=0.5,1,5$. Continuous lines represent curves of constant $t$ whereas dotted lines represent curves of constant $x$.}
\end{figure*}

In order to obtain a better understanding of the nature of the slices, we construct the conformal diagrams of the three cases studied. The conformal diagram is constructed using the coordinates $\tilde{t}=\tilde{t}(t,x)$ and $\tilde{x}=\tilde{x}(t,x)$. Considering that $h(\tilde{x}) = \sqrt{a^2 + \tilde{x}^2}$ and $\Omega = 1-x^2$, the physical spatial coordinate reads $\tilde{x}=\frac{x}{1-x^2}$. With this information one has the tilde coordinates given by:

\begin{equation}
\tilde{t} = \sqrt{a^2 + \frac{x^2}{(1-x^2)^2}}, ~~
\tilde{x} = \frac{x}{1-x^2}. \nonumber
\end{equation}

\noindent The results are depicted in Fig. \ref{fig:conformal_Mink1p1}. These diagrams are useful to interpret the effects of the gauge used. For the case of small $\tilde{k}$ the hypersurfaces used for the evolution, even if they reach $\scri^{+}$ they are concentrated near spatial infinity, whereas in the cases  of bigger values of $\tilde{k}$ slices are very separated near $\scri^{+}$ and allow the pulses arrive at the boundaries quickly.


\section{Schwarzschild space-time}


\subsection{Foliation and scri fixing}

We want to solve the wave equation on top of the Schwarzschild background in coordinates with two important properties: i) the wave is allowed to enter the event horizon of the black hole and ii) the wave is allowed to reach future null infinity. We then start with the Schwarzschild metric written as usual

\begin{equation}
d\tilde{s}^2 = -\left(1-\frac{2M}{\tilde{r}}\right)d\tilde{t}^2 + \frac{d\tilde{r}^2}{1-\frac{2M}{\tilde{r}}} +
	\tilde{r}^2 (d\theta^2 + \sin^2 \theta d\varphi^2),
\label{eq:sch_sch}
\end{equation}

\noindent and proceed to construct a hyperboloidal foliation and a scri-fixing compactification.

By introducing the change of coordinate $t=\tilde{t} - h(\tilde{r})$ in (\ref{eq:sch_sch}) the line element takes the form

\begin{eqnarray}\label{eq:slicing}
\nonumber d\tilde{s}^{2}&=& -\left(1-\frac{2M}{\tilde{r}}\right)dt^{2} - 2h'\left(1-\frac{2M}{\tilde{r}}\right)dtd\tilde{r} \nonumber\\ 
&+& \left[\left(1-\frac{2M}{\tilde{r}}\right)^{-1} - \left(1-\frac{2M}{\tilde{r}}\right)h^{'2} \right]d\tilde{r}^{2} \nonumber\\ 
&+& \qquad \tilde{r}^{2}(d\theta^{2} + \sin^2 \theta d\varphi^{2}), 
\end{eqnarray}

\noindent where $h^{\prime}=\frac{dh}{d\tilde{r}}$, and from which we read off the gauge quantities

\begin{eqnarray}
\tilde \alpha &=& \frac{1}{\tilde \gamma},\nonumber\\
\tilde{\beta} &=& -\frac{h'(1-\frac{2M}{\tilde{r}})}{\tilde{\gamma}^{2}},\nonumber\\
\tilde{\gamma}^{2} &=& \left(1-\frac{2M}{\tilde{r}}\right)^{-1} - \left(1-\frac{2M}{\tilde{r}}\right)h^{'2}.
\label{eq:gauge_sch}
\end{eqnarray}

\noindent from which the unit normal vector to the spatial hypersurfaces pointing to the future is given by

\begin{equation}
n^{\mu} = \left[ \tilde \gamma, \frac{h'}{\tilde \gamma}\left(1-\frac{2M}{\tilde{r}}\right), 0 ,0 \right].
\end{equation}


\noindent Now, the extrinsic mean curvature at the initial time slice $\tilde k = \tilde \nabla_{\mu} n^{\mu}$ reads

\begin{equation}
\tilde{k} = \frac{1}{\tilde{r}^{2}}\partial_{\tilde{r}}\left[\frac{\tilde{r}^{2}h'}{\gamma}\left(1-\frac{2M}{\tilde{r}}\right)\right],
\end{equation}

\noindent and can be integrated for constant $\tilde{k}$:

\begin{equation}
\frac{\tilde{k}\tilde{r}^{3}}{3}-C = \frac{\tilde{r}^{2}h'}{\gamma}\left(1-\frac{2M}{\tilde{r}}\right),
\end{equation}

\noindent where de derivative $h^{\prime}$ is

\begin{equation} \label{eq:h'}
h' = \frac{\left(\frac{\tilde{k}\tilde{r}^{3}}{3}-C \right)}{\left(1-\frac{2M}{\tilde{r}}\right)
\sqrt{\left(\frac{\tilde{k}\tilde{r}^{3}}{3}-C\right)^{2} + \left(1-\frac{2M}{\tilde{r}}\right)\tilde{r}^{4}}}. 
\end{equation}

\noindent In this case, it is not easy to find $h$ in a closed form as in the previous case in this paper, and in order to have a description of the slices one has to integrate this function numerically, however notice that $h$ is not required but only its derivative.

On the other hand, in order to perform the scri-fixing compactification we choose the conformal factor to be $\Omega=1-r$, and thus the compactifying coordinate is given by $\tilde{r}=\frac{r}{\Omega}$. Then the Schwarzschild space-time using scri-fixing conformal compactification is described by the line element \cite{Murc}

\begin{eqnarray}
ds^{2}&=&-\left(1-\frac{2M\Omega}{r}\right)\Omega^{2}dt^{2}-\frac{2(\tilde{k}r^{3}/3 -C\Omega^3)}{P(r)}dtdr \nonumber\\ 
&+& \frac{r^{4}}{P^{2}(r)}dr^2 + r^{2}(d\theta^{2} + \sin^2 \theta d\varphi^{2}),
\label{eq:Sch_compactified}
\end{eqnarray} 

\noindent where 
\begin{equation}
P(r)=\Omega^{3}\tilde{P}(r)= \sqrt{\left(\frac{\tilde{k}r^{3}}{3}-C\Omega^{3}\right)^{2} + \left(1-\frac{2M\Omega}{r}\right)\Omega^{2}r^{4}}.
\end{equation}

\noindent The values of $\tilde{k}$ and $C$ are restricted in such a way that $P(r)$ is real. 
Important features of these coordinates are that i) the event horizon is located at $r=2/3$, ii) the slices penetrate the event horizon, iii) the slices do not avoid the singularity at $r=\tilde r = 0$, and iv) they reach the future null infinity. 
Condition (ii) allows the practice of excision inside the horizon, that is, a chunk of the domain is removed from the numerical domain which we know lies inside the horizon of the black hole, which means that  the domain $r \le r_{exc}$ is removed and no need of boundary conditions at $r_{exc}$ are needed as long as the speeds of the characteristic fields are negative at that boundary.


The gauge functions for this conformal metric are in this case 

\begin{eqnarray}
\alpha &=& \sqrt{\left(1- \frac{2\Omega}{x}\right) \Omega^2 + \frac{(C\Omega^3 - \frac{1}{3}\tilde k x^3)^2}{x^4}},\nonumber\\
\beta &=& \frac{P(r)}{x^4} \left( C\Omega^3 - \frac{1}{3}\tilde k x^3 \right),\nonumber\\
\gamma &=& \frac{x^2}{P(r)} \label{eq:GaugeSchw}
\end{eqnarray}

\subsection{The wave equation}

Following our convention, the conformal metric is given by $ds^2 = \Omega^2 d\tilde{s}^2$, meaning that the physical metric is the one with the tildes. We decide to solve the wave equation using the conformal metric because of various reasons: 1) the slices are hyperboloidal and reach $\scri^{+}$ at infinity, 2) the spatial coordinate is compactified, and therefore the wave function reaches future null infinity at the boundary $r=1$, 3) such boundary is null and there is no need to impose boundary conditions there, because the characteristic speed $\lambda_{-} = 0$ on that boundary.
In order to use these benefits it is needed to solve the conformally invariant wave equation

\begin{equation}
\left( \Box - \frac{1}{6}R \right)\phi_T(t,r,\theta,\varphi) = 0,
\label{eq:conf_inv}
\end{equation}

\noindent where $R$ is the Ricci scalar of the conformal metric and $\Box=\nabla^{\mu}\nabla_{\mu}$ corresponds also to the conformal metric. In order to study non-radial modes, we separate $\phi_T(t,r,\theta,\varphi)=\phi(t,r) Y_{lm}(\theta,\varphi)$ with $Y_{lm}$ the spherical harmonics.  We solve numerically this equation considering a domain $r \in [r_{exc},1]$. We choose $r_{exc}$ such that it lies inside the event horizon and satisfies the need of $P(r)$ being real. 

We solve (\ref{eq:conf_inv}) for $\Phi(t,r,\theta,\varphi)$ as an initial value problem using a first order variable formulation. In terms of the line element given by the gauge (\ref{eq:GaugeSchw}) the system of equations to be solved is

\begin{eqnarray}
\partial_t \psi&=&  \partial_r\left( \frac{\alpha}{\gamma}\pi + \beta \psi \right),\label{eq:wavetail}\\
\partial_t \pi&=& \frac{1}{r^2} \partial_r \left(r^2 (\beta \pi + \frac{\alpha}{\gamma}\psi ) \right)
- \alpha \gamma \left( \frac{1}{6}R + \frac{l(l+1)}{r^2} \right)\phi, \nonumber\\
\partial_r \phi&=&\psi, \nonumber
\end{eqnarray}

\noindent where $R = \frac{12 \Omega}{r^2}\left( r + m(2r-1) \right)$. The system of equations is again symmetric hyperbolic and we solve it using initial data for $\pi$ and $\psi$ in the domain $r \in [r_{exc},1]$ and $t \in [0,\infty)$. As initial data we choose an approximately outgoing Gaussian pulse that on the initial hypersurface satisfies

\begin{equation}
\partial_{t} (r \phi) + \lambda_{+}\partial_{r} (r \phi) =0,  \label{eq:outgoing}
\end{equation}

\noindent where $\lambda_{+}$ is the outgoing characteristic speed given by $\lambda_{+}= -\beta +\frac{\alpha}{\gamma}$. The initial data reads as follow

\begin{eqnarray}\label{eq:evoldata}
\phi(0,r) &=&  Ae^{-(r-r_{0})^2/\sigma^2}, \nonumber \\
\psi(0,r) &=& -2 \frac{(r-r_{0})}{\sigma^2}\phi(0,r),\\
\pi (0,r) &=& -\psi(0,r) - \frac{\phi(0,r)}{r}\left( 1 - \frac{\beta\gamma}{\alpha}\right). \nonumber
\end{eqnarray}

The numerical methods consist again in a finite differences approximation using the method of lines, however this time we use more accurate stencils in order to track small amplitudes related to the tail decay behavior of the wave function. We use now sixth order stencils along the spatial direction and a fourth order Runge-Kutta integrator in time.

We choose an excision radius $r_{exc}<2/3$ such that in the numerical domain $P(r)$ is real for the values of $C$ and $\tilde{k}$ we have used. About the boundary at future null infinity $r=1$, the characteristic speed of the mode moving to the left $\lambda_{-} = -\beta - \alpha / \gamma$ is zero, so that the wave propagates along the boundary and not toward the numerical domain. This implies that there is no need to apply boundary conditions there.

\begin{figure*}[ht]
\includegraphics[width=8.5cm]{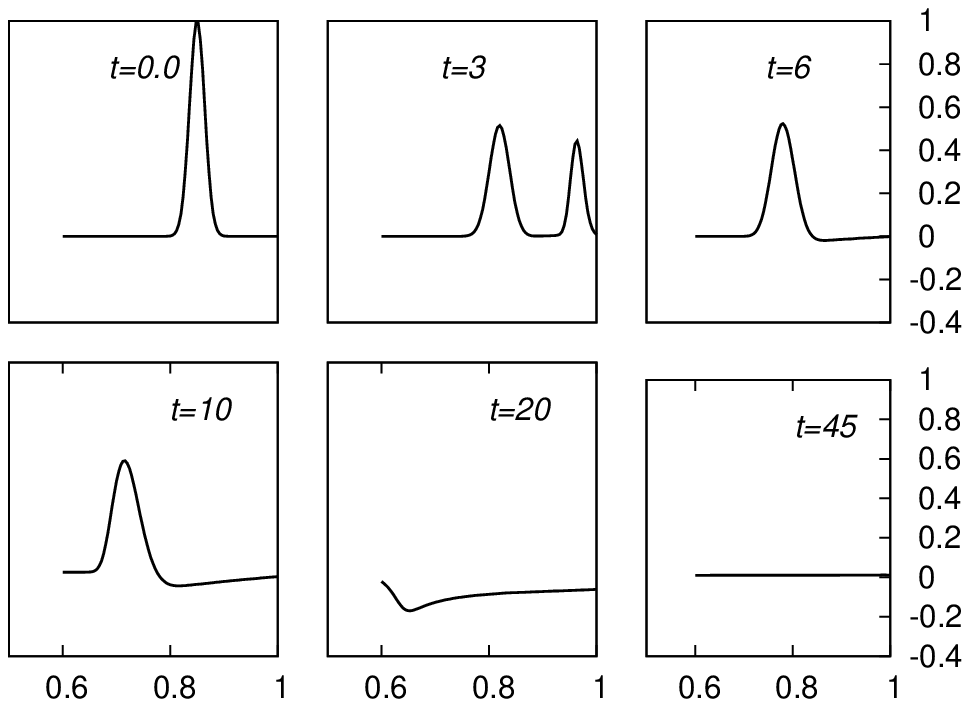}
\includegraphics[width=8.5cm]{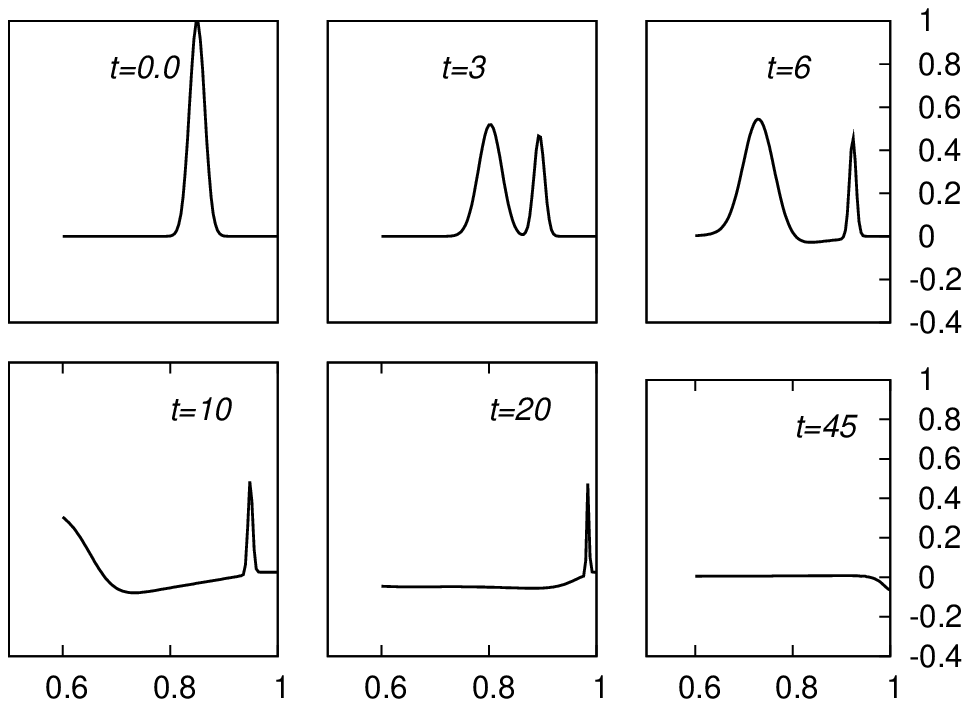}
\caption{\label{fig:Sch_compact_snaps} Snapshots of the wave function for $\tilde{k}=0.4, 0.1$, left and right panels respectively for the wave equation on the space-time given by (\ref{eq:Sch_compactified}). Again, slices with small curvature imply the outgoing pulse squeezes and slows down before it reaches $\scri^{+}$, whereas for big curvature values the outgoing pulse reaches future null infinity quickly. The initial Gaussian is located at $r=0.8$. The pulse splits into two pulses one moving toward scri plus and the other moving toward the horizon. We excise a part of the domain with $r_{exc}=0.6$.}
\end{figure*}

\begin{figure*}[ht]
\includegraphics[width=7cm]{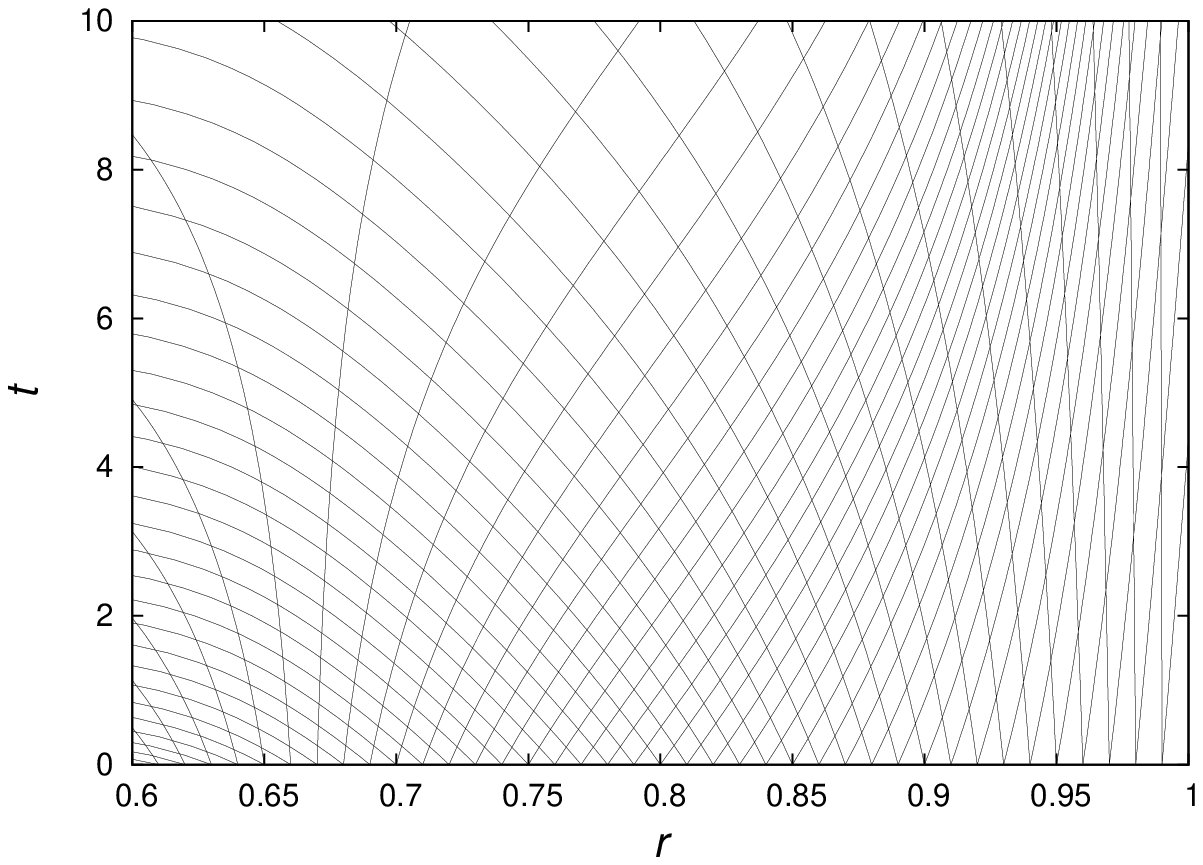}
\includegraphics[width=7cm]{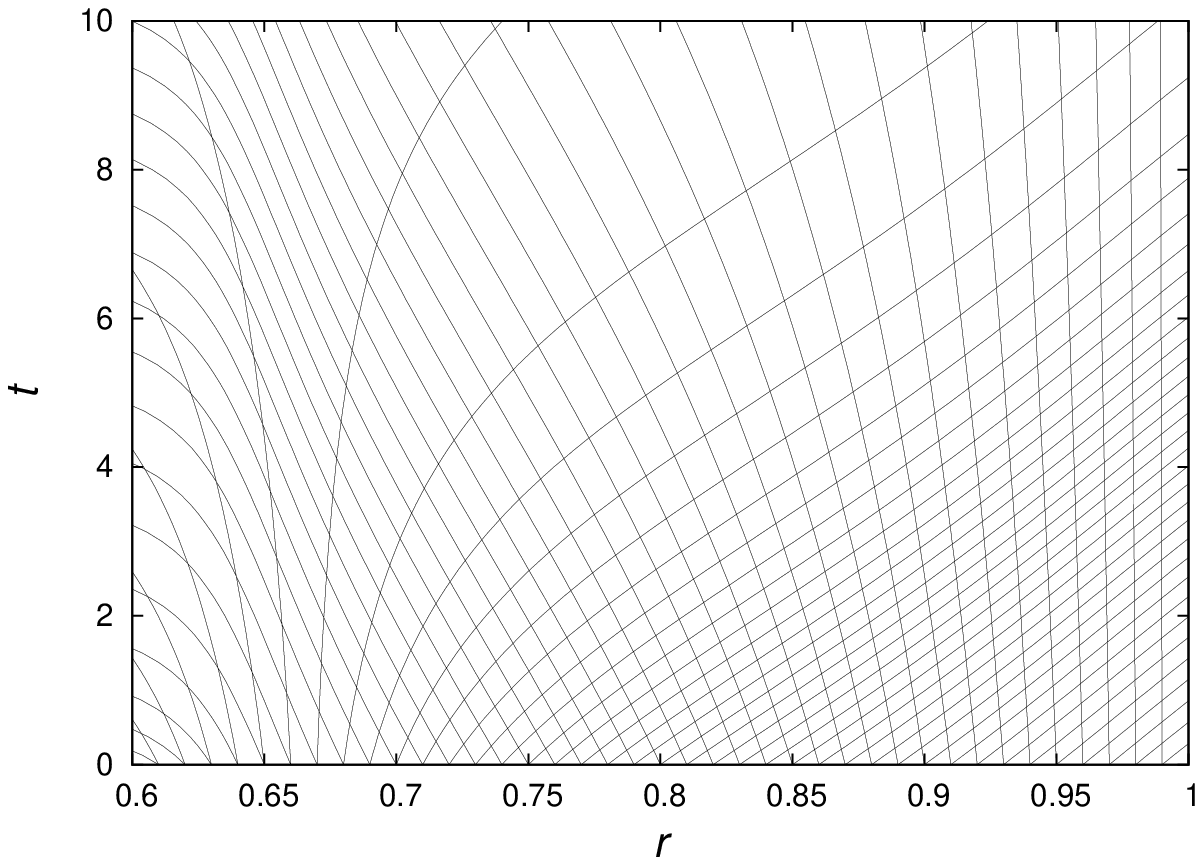}
\caption{\label{fig:Sch_cones} We show the light cones for the space-time (\ref{eq:Sch_compactified}) for the two values $\tilde{k}=0.1$ (left) $\tilde{k}=0.4$ (right). 
It can be seen that for small $\tilde{k}$ the light cones stretch near the boundary whereas for big values of $\tilde k$ the cones are wide open at $\scri^{+}$.}
\end{figure*}

In Fig.\ref{fig:Sch_compact_snaps} we show the effects of the gauge parameters chosen for the slices of the space-time for a spherical wave. In one case, when a large value of the curvature is used, the light cones near $\scri^{+}$ are wide open and the outgoing pulse of scalar field arrives at such boundary without suffering a considerable deformation, whereas for smaller values of the curvature the pulse slows down and squeezes a little. To explain this effects we show in Fig. \ref{fig:Sch_cones} the light cone structure of the space-time for two values of the extrinsic curvature of the slices.
An important comment about the location of event horizons from Fig. \ref{fig:Sch_cones} is also in turn. Event horizons are located in numerical relativity via tracking null surfaces, and the event horizon is calculated approximately as the surface from which the evolution of null 2-surfaces to the future diverge toward inside the horizon and toward $\scri^{+}$ \cite{diener}. In this case, where hyperboloidal slices are compactified and the causal structure is maintained even after the conformal rescaling, we are sure that some of the null rays reach $\scri^{+}$ located at $r=1$; a fine tuned set of null rays in Fig. \ref{fig:Sch_cones} would locate the exact threshold between those outgoing rays (rays in this case of spherical symmetry) really escaping to future null infinity and those that reach the excision boundary, being such threshold the location of the event horizon at $r=2/3$. 


\subsection{Tails decay rates and quasinormal modes for the test scalar field on Schwarzschild space-time}

The study of wave tails in general relativity was first studied by Price and Pullin \cite{price,pullin}.
Polynomial tail decay shows up after an oscillatory phase related to the quasinormal ring-down frequencies of black holes and can be regarded as being due to back-scattering of gravitational radiation off the curved background space-time. It is estimated that polynomial tail decay $\sim t^p$ depends on the mode $l$ that is being analyzed and on the trajectory of the observer, for instance $p=-3$ for time-like observers \cite{price} whereas for null observers at $\scri^{+}$ $p=-2$ \cite{pullin} when $\l=0$, and in general $p= -2l-3$ at finite distances and $p=-l-2$ at $\scri^{+}$. 

Recent studies about the tail decay of a mass-less scalar field have been presented, which include the dependence on the location of the observer \cite{anil2}. In this section we reproduce the decay rates and quasinormal frequencies for a mass-less scalar field with angular momentum on a fixed Schwarzschild space-time background for $l=0,1,2$.  In order to do so, we solve again the initial value problem for (\ref{eq:wavetail},\ref{eq:outgoing},\ref{eq:evoldata}) and study the amplitude of $\phi$ for large values of time.

\subsection{Numerical results}

In order to study the oscillations and the tail decay it is enough to track the behavior in time of the amplitude of the physical scalar field $\tilde \phi$, which reveals the oscillations and polynomial tail decay. Even though we are solving for the test scalar field on top of the conformal metric, the amplitude of the conformal scalar field $\phi$ will show the same behavior in time as the physical scalar field, this is why showing the results for $\phi$ suffices, since the conformal factor depends only on the spatial coordinate. We locate detectors of the scalar field at various positions in the domain, and relate the physical and conformal positions via the definition of the compact coordinate $\tilde r= r/(1-r)$ and its inverse $r = \tilde r / (1+\tilde r)$.

\begin{figure}[ht]
\includegraphics[width=8cm]{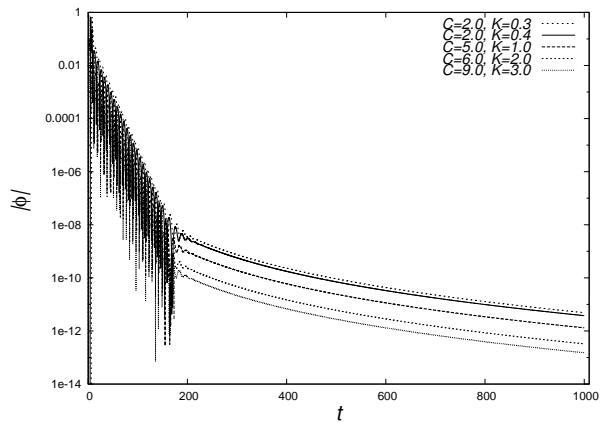}
\caption{\label{fig:kc} We show the amplitude of the test scalar field in time, including the oscillating and the polynomial tail decay stages for different values of $C$ and $K$ measured at $\scri^{+}$ on Schwarzschild space-time for $l=2$ and $M_{ADM}=1$. We found that each observer measures the same decay exponent $p$ for different values of $C$ and $K$ and also the quasinormal mode frequencies measured are the same for different values of $C$ and $K$.}
\end{figure}

The typical behavior of the wave function amplitude at late times is shown in Fig. \ref{fig:kc} for various values of the gauge parameters. At initial times we have the quasinormal mode oscillations, after which the polynomial tail decay appears. We have also verified that the exponents are independent of a wide range of initial data profiles, both, amplitudes and widths of the Gaussian.

\begin{figure}[ht]
\includegraphics[width=8.5cm]{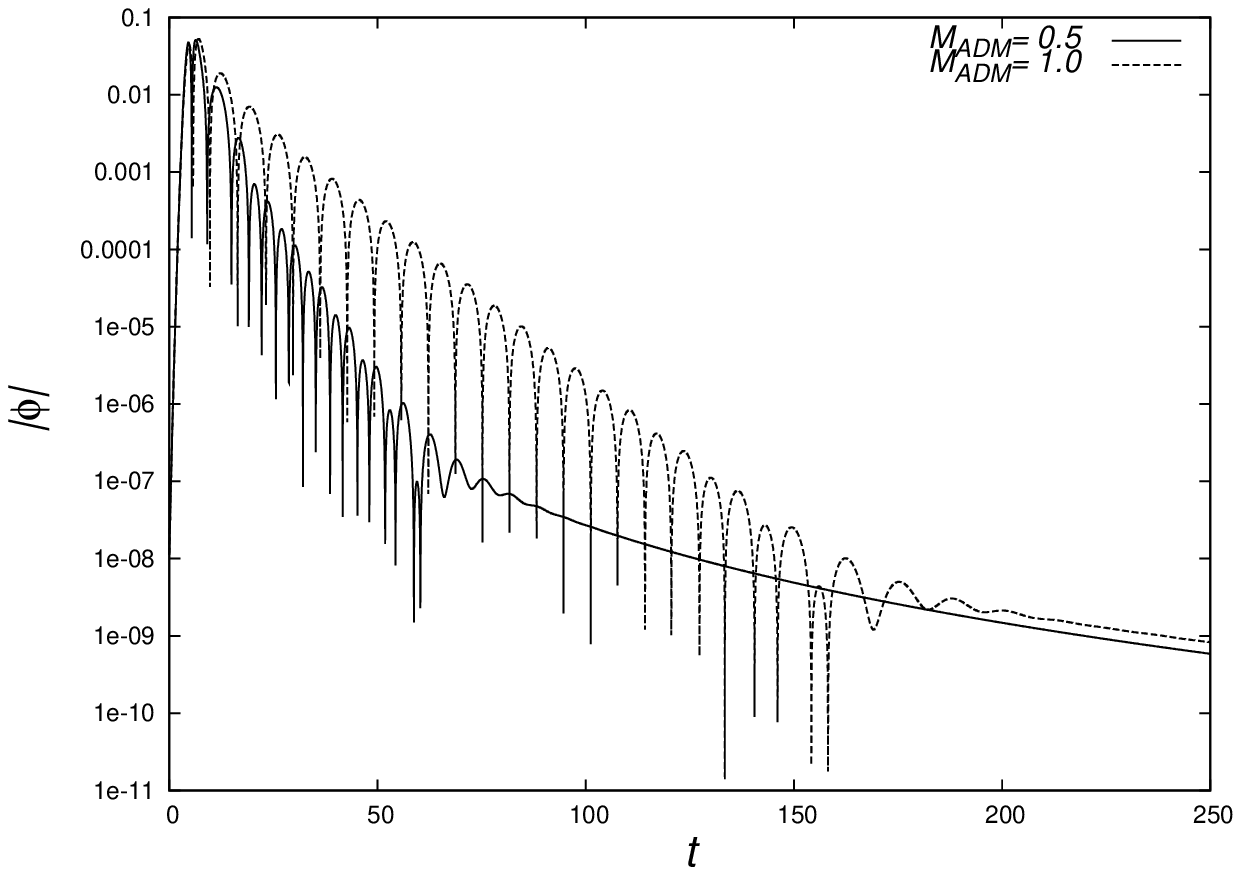}
\includegraphics[width=8.5cm]{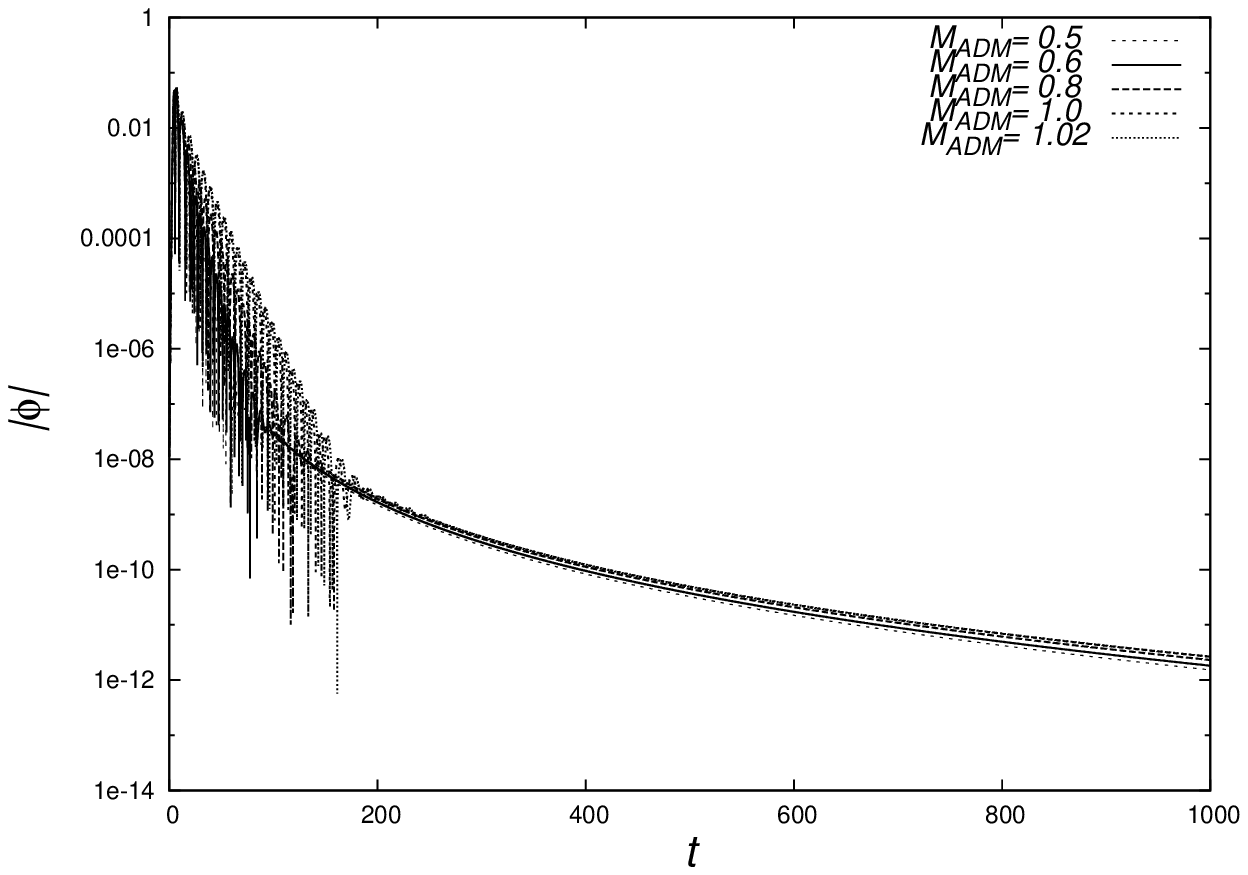}
\caption{\label{fig:admmass} We show the quasinormal modes and the tail decay rates of the wave function for different values of the mass of the black hole, measured an observer located at $\scri^{+}$ for $l=2$, $K=0.4$ and $C=2$. We found that the frequencies are different for differents ADM masses  and the decay exponents are the same. The figure in the left panel is a zoom of the plot in the right panel for two cases that shows more clearly the dependence of the frequency on the mass of the black hole. It can be noticed also that the tail decay exponent is the same in all cases.}
\end{figure}

An important property that can be explored at this point is the frequency of the quasinormal mode oscillations. In Fig. \ref{fig:admmass} we show the amplitude of the scalar field for two different values of the ADM mass $M$, in one case we use $M=0.5$ as usually done in previous papers, and in the other case we use the value $M=1$ which is the one we use in all our further calculations. A fit of the oscillations shown in Fig. \ref{fig:admmass} with the function $Ae^{i(\omega+\omega_I)t}$ reveals that the frequencies for the mode $l=2$, using $M=0.5$ are $\omega = 0.749$ and $\omega_{I} = -0.178$ which is consistent with previous studies \cite{Leaver,kokkotas}. 

\begin{figure}[ht]
\includegraphics[width=8cm]{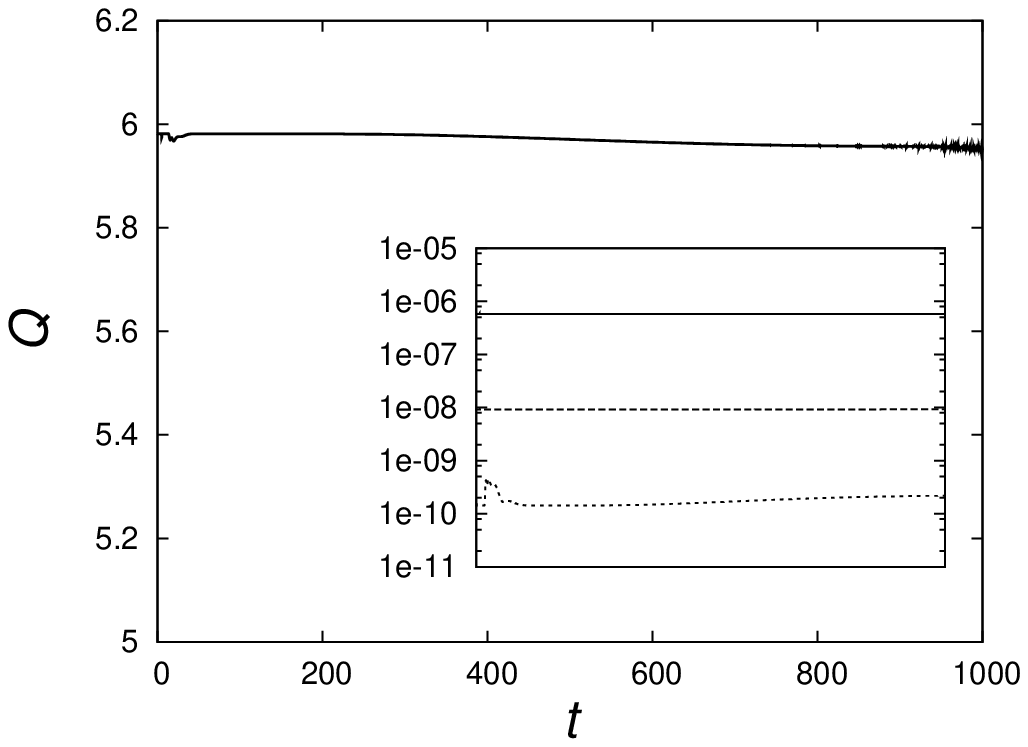}
\includegraphics[width=8.5cm]{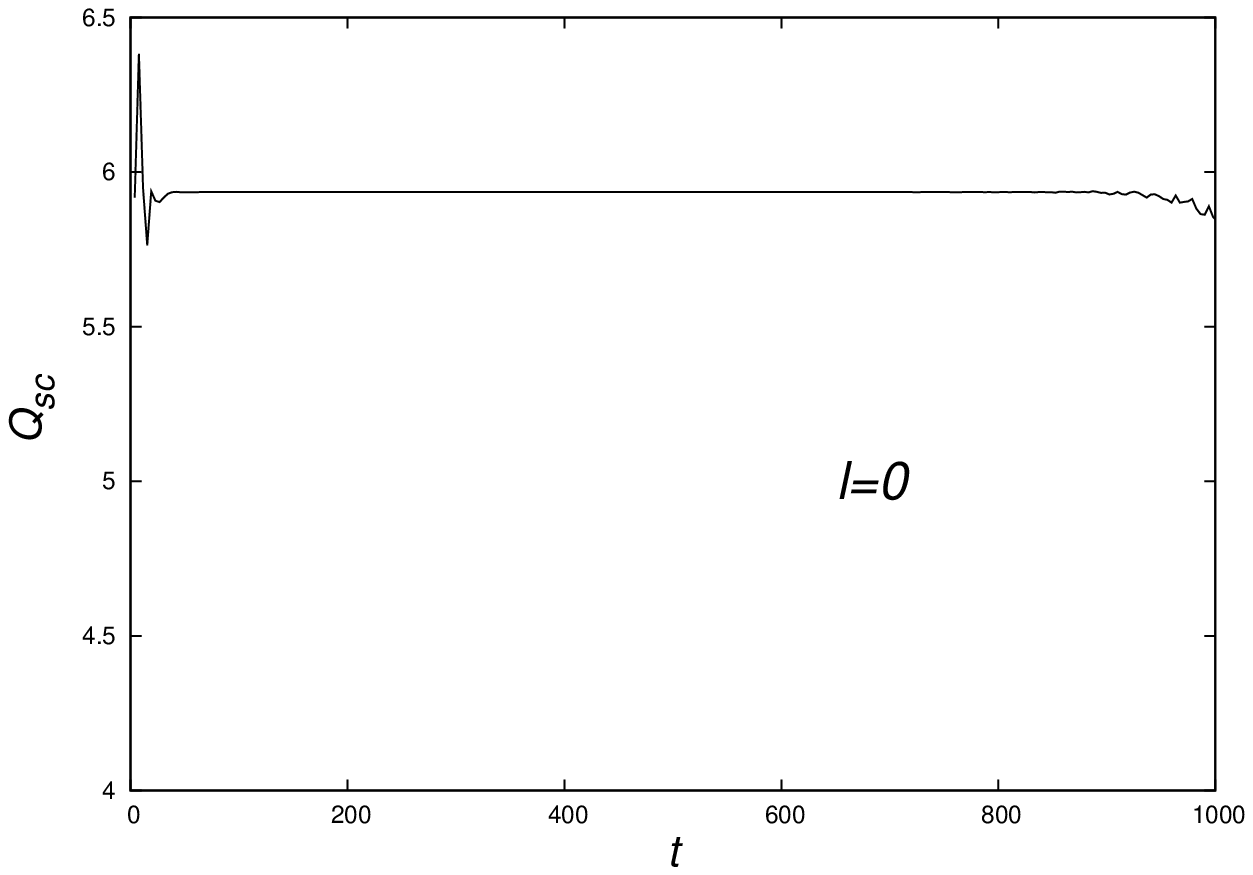}
\caption{\label{fig:convergence} We show the typical convergence tests for one of our runs, in this case for $l=0$ and resolutions  $\Delta x_1 = 7.6 \times 10^{-4}$, $\Delta x_2 = \Delta x_1/2$, $\Delta x_3 = \Delta x_1/4$. In the left panel we show the order of convergence $Q$ of the $L_2$ norm of the constraint $C=\psi-\partial_r \phi$ as defined in Fig 2, and in the inset the $L_2$ norm of the violation of the constraint for the three resolutions. 
In the right panel we show the order of self-convergence $Q_{sc}$ of $\phi$, using the $L_2$ norm of the differences between the value of the wave function for the various resolutions.}
\end{figure}

In order to validate our numerical results we show in Fig. \ref{fig:convergence} an example of the convergence of the constraint $\psi=\partial_r \phi$ and snapshots of the self convergence tests of the scalar field. These tests indicate that the convergence of our results range between fourth and sixth order for quite a long time, which is consistent with the order of accuracy of our finite difference approximation and our MoL time-integrator.

\begin{figure}[ht]
\includegraphics[width=8.5cm]{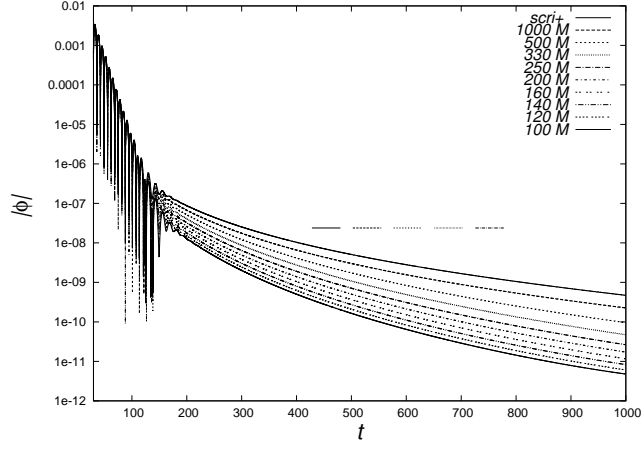}
\caption{\label{fig:tail_dep_det} Amplitude of the wave function measured at different distances from the black hole's horizon for $l=2$, which indicates that the tail decay exponent depends on the location of the detector.  We use: $A =0.1$, $x_{0}=0.8$ and $\sigma=0.1$.}
\end{figure}

We also measure the wave function for observers located at different positions from the event horizon. In Fig. \ref{fig:tail_dep_det} we show how the polynomial tail decay exponent depends on the location of the detector for $l=2$, and similar behavior is found for the other modes.

\begin{figure}[ht]
\includegraphics[width=5.5cm]{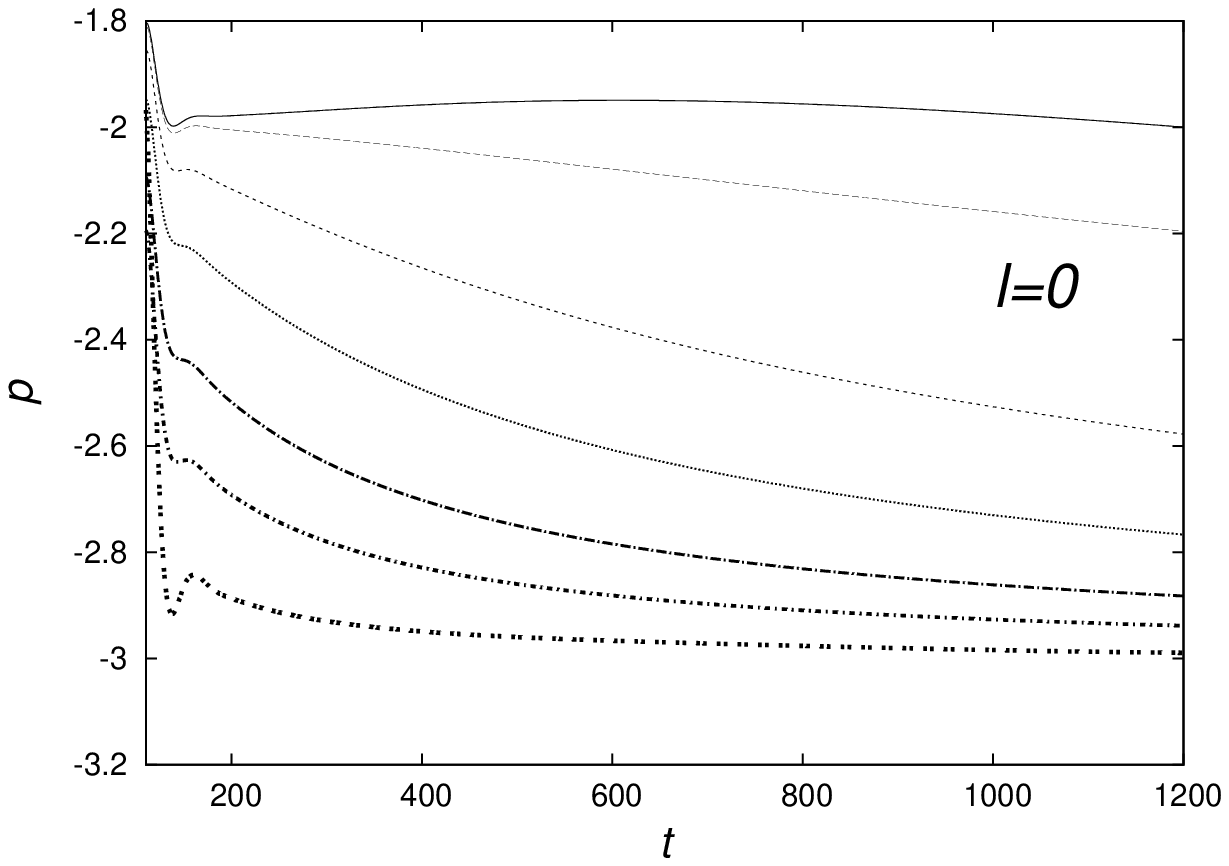}
\includegraphics[width=5.5cm]{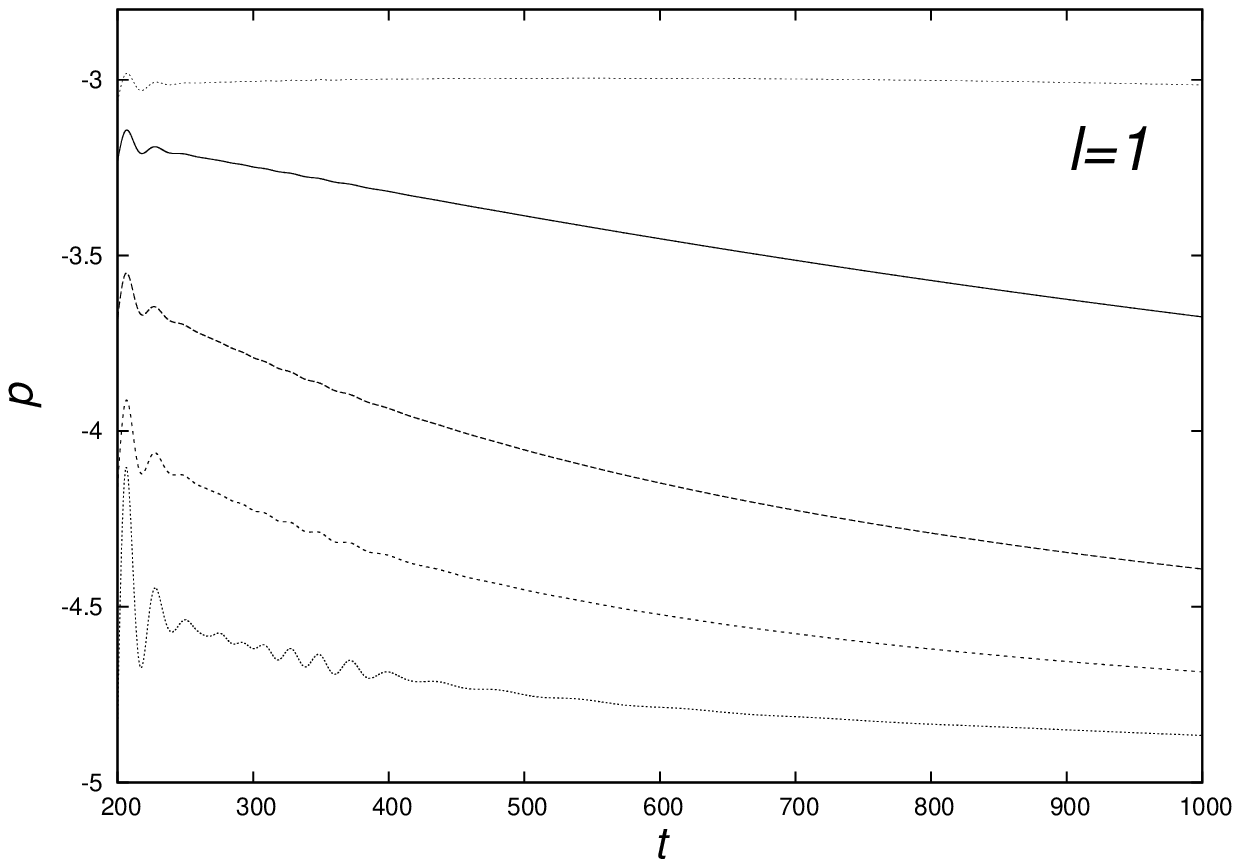}
\includegraphics[width=5.5cm]{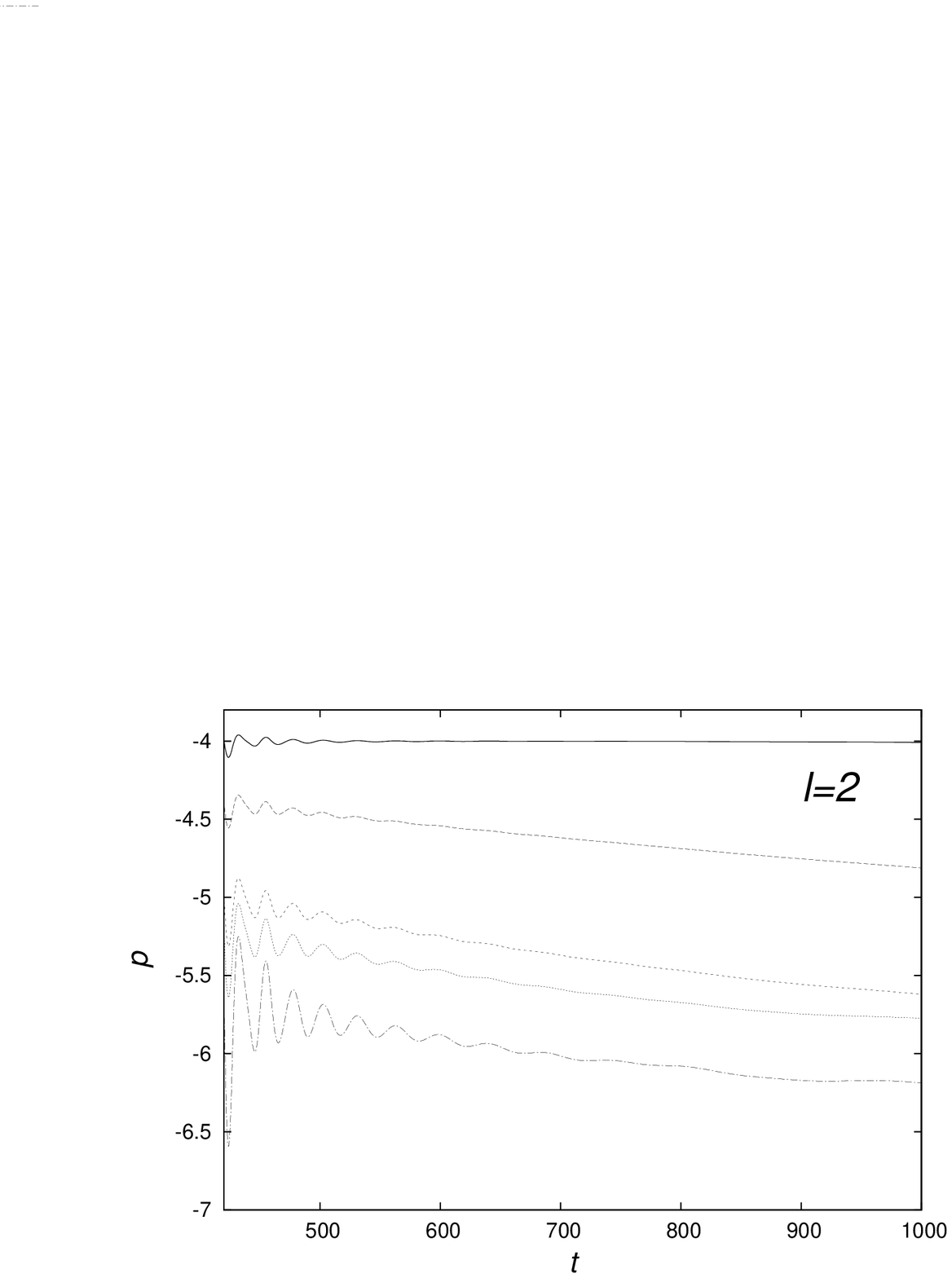}
\caption{\label{fig:exponents} We show the exponent of the polynomial tail decay for $l=0,1,2$ measured by observers located at various distances from the black hole. From top to bottom the curves correspond to detectors located at $\tilde r= \infty,~ 1000M,~ 250M,~ 110M,~ 50M,~ 10M$ and the remaining curves (for the case $l=0$) correspond to detectors located even closer to the horizon.}
\end{figure}

An efficient way of calculating the exponent $p$ consists in using the fact that assuming $\phi(x,t) \sim K t^p$ with $K$ a constant, the exponent is given by $p = \frac{d \log |\phi|}{d \log (t)}$ as done in \cite{anil2}, so that it is possible to track the time dependence of the exponent itself. In Fig. \ref{fig:exponents} we show the exponent $p$ for $l=0,1,2$, measured at various positions and verify the bounds of the exponents to be between $p=-2l-3$ for time-like observers and $p=-l-2$ at $\scri^{+}$. 

\section{General formulas for the scri-fixing conformal compactification}

We now develop the construction of the conformal compactification for a static spherically symmetric space-time described by the type of metric 

\begin{equation}
d\tilde{s}^2 = -\alpha^2(\tilde{r}) d\tilde{t}^2 + b^2(\tilde{r}) d\tilde{r}^2 +
	\tilde{r}^2 (d\theta^2 + \sin^2 \theta d\varphi^2),
\label{eq:Boson-metric}
\end{equation}

\noindent where $b$ and $\alpha$  are assumed to be known metric functions, and proceed to construct a hyperboloidal foliation and a scri-fixing conformal compactification. By introducing the change of coordinate $t=\tilde{t} - h(\tilde{r})$  the line element takes the form

\begin{eqnarray}\label{eq:slicing}
\nonumber && d\tilde{s}^{2}=-\alpha^2(\tilde{r})dt^{2} - 2h'(\tilde{r})\alpha^2(\tilde{r})dtd\tilde{r} \nonumber\\ 
&& + \left[b^2(\tilde{r})  - \alpha^2(\tilde{r}) (h'(\tilde{r}))^2 \right]d\tilde{r}^{2} +  \tilde{r}^{2}(d\theta^{2} + \sin^2 \theta d\phi^{2}), 
\end{eqnarray}

\noindent where $h^{\prime}=\frac{dh}{d\tilde{r}}$. Comparison of this metric with the 3+1 metric 
$d\bar{s}^2 = (-\bar{\alpha}^2 + \bar{\gamma}^2 \bar{\beta}^2)dt^2 + 2 \bar{\beta}\bar{\gamma}^2 d \tilde{r}dt + \bar{\gamma}^2 d\tilde{r}^2 + \tilde{r}^2 (d\theta^2 + \sin^2 \theta d\varphi^2)$ allows to read off the gauge and metric functions

\begin{eqnarray}
\bar{\alpha}(\tilde{r}) &=& \frac{\alpha(\tilde{r}) b(\tilde{r})}{\bar{\gamma}(\tilde{r})},\nonumber\\
\bar{\beta}(\tilde{r}) &=& -\frac{h'(\tilde{r})\alpha^2(\tilde{r})}{\bar{\gamma}^{2}(\tilde{r})},\nonumber\\
\bar{\gamma}^{2}(\tilde{r}) &=& b^2(\tilde{r}) - \alpha^2 (\tilde{r})(h'(\tilde{r}))^2.
\label{eq:gauge_sch}
\end{eqnarray}

In terms of these functions, the unit normal vector to the spatial hypersurfaces pointing to the future is given by

\begin{equation}
n^{\mu} = \left[ \frac{\bar{\gamma}(\tilde{r})}{\alpha(\tilde{r})b(\tilde{r})}, \frac{h'(\tilde{r})\alpha(\tilde{r})}{\bar{\gamma}(\tilde{r})b(\tilde{r})}, 0 ,0 \right].
\end{equation}

Given a space-like slice, we can drag it along the time-like Killing vector. This  will give a slicing where the time translation is along the Killing vector. Now, the mean curvature at the initial time slice, is given by

\begin{equation}
\tilde{k} = \frac{1}{\tilde{r}^{2}\alpha(\tilde{r})b(\tilde{r})}\partial_{\tilde{r}}\left[\frac{\tilde{r}^{2}h'(\tilde{r})\alpha^2 (\tilde{r})}{\bar{\gamma}(\tilde{r})}\right],
\end{equation}

\noindent and can be integrated for constant $\tilde{k}$, which means the slices are of constant mean curvature (CMC):

\begin{equation}
\tilde{k}\int \tilde{r}^2 \alpha(\tilde{r})b(\tilde{r})d\tilde{r} - C = \frac{\tilde{r}^{2}h'(\tilde{r})\alpha^2 (\tilde{r})}{\bar{\gamma}(\tilde{r})},
\end{equation}

\noindent where now $h'(\tilde{r})$ is

\begin{equation} \label{eq:h'}
h'(\tilde{r}) = \frac{[\tilde{k}I(\tilde{r}) - C]b(\tilde{r})}{\alpha(\tilde{r})\sqrt{[\tilde{k}I(\tilde{r}) - C]^2 + \alpha^2(\tilde{r})\tilde{r}^4}}
\end{equation}

\noindent and 

\begin{equation}
I(\tilde{r}) = \int \tilde{r}^2 \alpha(\tilde{r})b(\tilde{r})d\tilde{r}.
\label{eq:iintegral}
\end{equation} 

\noindent In general, it is not easy to find $h$ in a closed form, so in order to have a description of the slices in general one has to integrate this function numerically. 

On the other hand, in order to perform the scri-fixing compactification,  we define the compact coordinate $r$ by $\tilde{r}=\frac{r}{\Omega}$ and we rescale the original metric using the conformal factor $\Omega$. The space-time using scri-fixing conformal compactification is finally given by the conformal line element 

\begin{eqnarray}
\nonumber && ds^{2}=-\alpha^2(r)\Omega^2(r) dt^{2} - 2h'(r)\alpha^2(r)(\Omega - r\Omega')dtdr \nonumber\\ 
&& + \left[b^2(r)  - \alpha^2(r) (h'(r))^2 \right](\Omega-r\Omega')^2\frac{dr^{2}}{\Omega^2(r)}  \nonumber \\ 
&& +  r^{2}(d\theta^{2} + \sin^2 \theta d\varphi^{2}), 
\label{eq:Sph_compactifiedgeneral}
\end{eqnarray} 
 
\noindent where the functions $\alpha(r)$, $b(r)$ and $h'(r)$  are functions of $r$. The conformal factor $\Omega$ determines various properties of the resulting conformal metric.

For instance, for the Schwarzschild space-time, $\alpha^2 = \frac{1}{b^2}= (1-\frac{2M}{\tilde{r}})$, with which our expression (\ref{eq:h'}), using $\Omega=1-r$ is reduced to the expression for $h'(\tilde{r})$ obtained by Malec and Murchadha \cite{Murc}, and the final version of the conformally rescaled metric is (\ref{eq:Sch_compactified}).


\subsection{The wave equation}

In general, from (\ref{eq:Sph_compactifiedgeneral}) we can read off again the gauge in terms of the ADM-like metric $d\hat{s}^2 = (-\hat{\alpha}^2 + \hat{\gamma}^2 \hat{\beta}^2)dt^2 + 2 \hat{\beta}\hat{\gamma}^2 drdt + \hat{\gamma}^2 dr^2 + r^2 (d\theta^2 + \sin^2 \theta d\varphi^2)$ and obtain the following gauge and metric functions

\begin{eqnarray}
\hat{\alpha}^{2}(r) &=& \alpha^2(r) \Omega^2 + \hat{\beta}^2(r) \hat{\gamma}^2(r),\nonumber\\
\hat{\beta}(r) &=& -\frac{\alpha^2(r) h'(r) (\Omega - r\Omega')}{\hat{\gamma}^2(r)},\nonumber\\
\hat{\gamma}^2(r) &=& \frac{(b^2(r) - \alpha^2(r) h'^2(r))(\Omega-r\Omega')^2}{\Omega^2},\nonumber\\
\label{eq:gauge_sch}
\end{eqnarray}

\noindent which are the final metric functions used to solve the wave equation for a generic space-time. In order to solve (\ref{eq:conf_inv}) as a first order system, first order variables are define as previously $\pi = \frac{\hat \gamma}{\hat \alpha}\partial_t \phi - \frac{\hat \gamma}{\hat \alpha} \hat \beta \partial_r \phi$ and $\psi = \partial_r \phi$. Then the system to be solved is (\ref{eq:wavetail}) with the gauge (\ref{eq:gauge_sch}) for given initial data on a given spatial domain and appropriate boundary conditions.



\subsection{ Example: 3+1 Minkowski space-time}

As an extra example, we present the case of the Minkowski space-time in spherical coordinates. We start with the line element for the Minkowski space-time in spherical coordinates $x^{\mu}=(\tilde{t},\tilde{r},\theta,\phi)$ for the physical metric given by

\begin{equation}
d\tilde{s}^2 = -d\tilde{t}^{2} + d\tilde{r}^{2} + \tilde{r}^2 (d\theta^2 + \sin^2 \theta d\phi^2),
\end{equation}

\noindent where $\tilde{t}\in(-\infty,\infty)$ and $\tilde{r}\in[0,\infty)$. The integral (\ref{eq:iintegral}) implies $I(\tilde r)= \tilde r^3 /3$, which substituted into (\ref{eq:h'}) with $C=0$ implies 

\begin{equation}
h'= \frac{\tilde{r}}{\sqrt{a^{2}+\tilde{r}^{2}}},
\end{equation}

\noindent where $a$ is a constant that is given in terms of the curvature by $\tilde{k}=\frac{3}{a}$. It is also possible to integrate again for $h$ and obtain a complete description of the slices with the height function $h(\tilde{x})= \sqrt{a^{2}+ \tilde{x}^{2}}$, which completes the description of the conformal metric of the space-time.


\subsection{A particular case of the 3+1 Minkowski space-time}

In order to compactify this case we use a compact coordinate $\tilde{r}=\frac{r}{\Omega}$. Unlike the 1+1 Minkowski case, one only needs to regularize at $r=1$ for which various choices are evident: $\Omega=1-r$ or $\Omega=1-r^2$, in fact in \cite{anilphi3} for the solution of the Klein-Gordon equation with a $\phi^4$ potential, the conformal factor used is $\Omega=(1-r^2)/2$, which is a conformal factor that shows a regular Ricci Scalar $R$ at the origin. For the example developed here we choose the first choice and the conformal metric then reads:

\begin{eqnarray}
ds^2 &=&  -(1-r)^2dt^2 - \frac{2r}{\sqrt{a^2 (1-r)^2 + r^2}}dtdr \nonumber\\
	&+& \frac{a^2}{a^2 (1-r)^2 + r^2}dr^2 + r^2(d\theta^2 + \sin^2 \theta d\phi^2),
	\label{eq:metric_mink3p1_conformal}
\end{eqnarray}

\noindent from which we read off the gauge functions 

\begin{eqnarray}
\alpha^2 &=& (1-r)^2 + \frac{r^2}{a^2},\nonumber\\
\beta   &=& -\frac{r}{a^2}\sqrt{a^2 (1-r)^2 + r^2}, \nonumber\\
\gamma^2 &=& \frac{a^2}{a^2 (1-r)^2 + r^2}.\label{eq:gauge_Mink3p1_wave}
\end{eqnarray}

In order to visualize the space-time structure we calculate the null rays of the space-time which obey $ds=0$, that is

\begin{equation}
\frac{dt}{dr} = \frac{1}{(1-r)^2}\left(- \frac{r}{\sqrt{a^2 (1-r)^2 + r^2}} \pm 1\right),
\end{equation}
\noindent whose solution reads

\begin{equation}
t -t_0= \frac{\sqrt{(a^2+1)(r-1)^2+2r-1}}{r-1}\pm\frac{1}{1-r}.
	\label{eq:mink_specific_example}
\end{equation}

\begin{figure*}[ht]
\includegraphics[width=5cm]{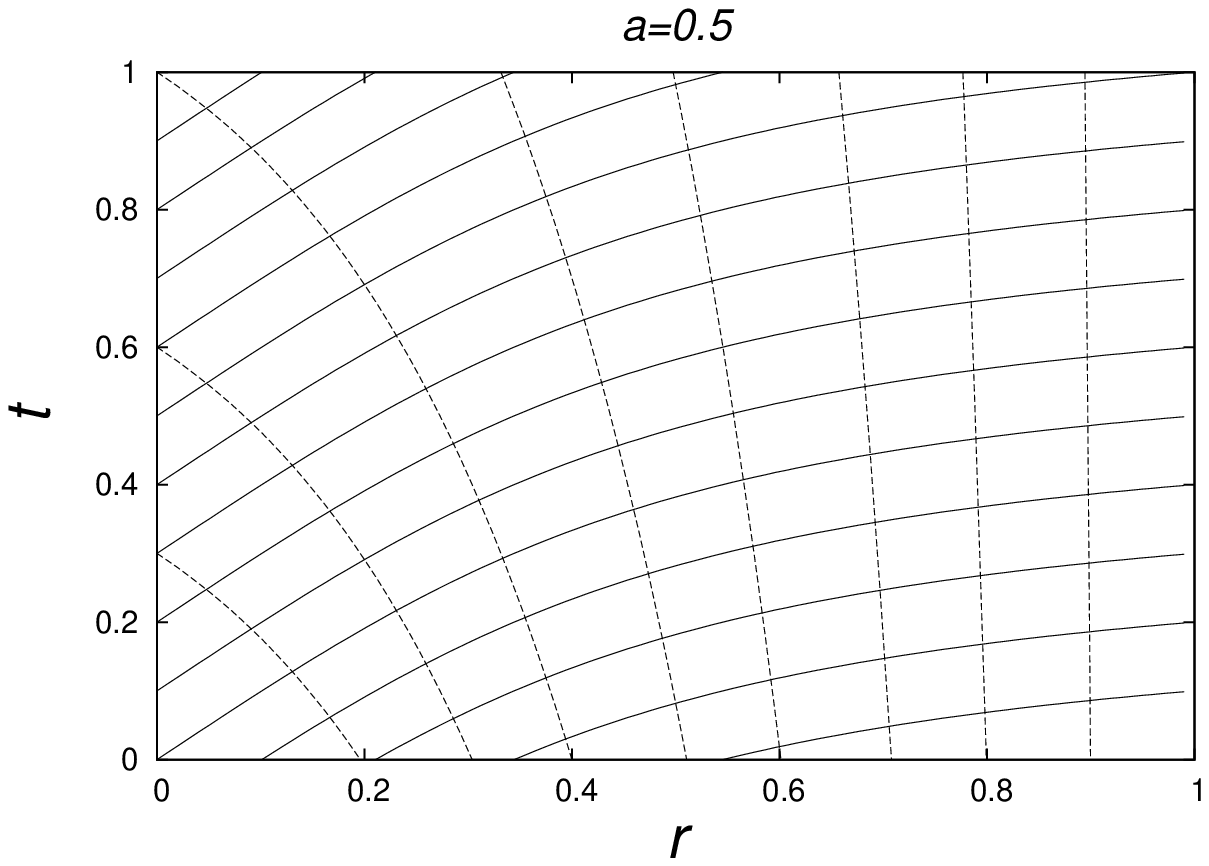}
\includegraphics[width=5cm]{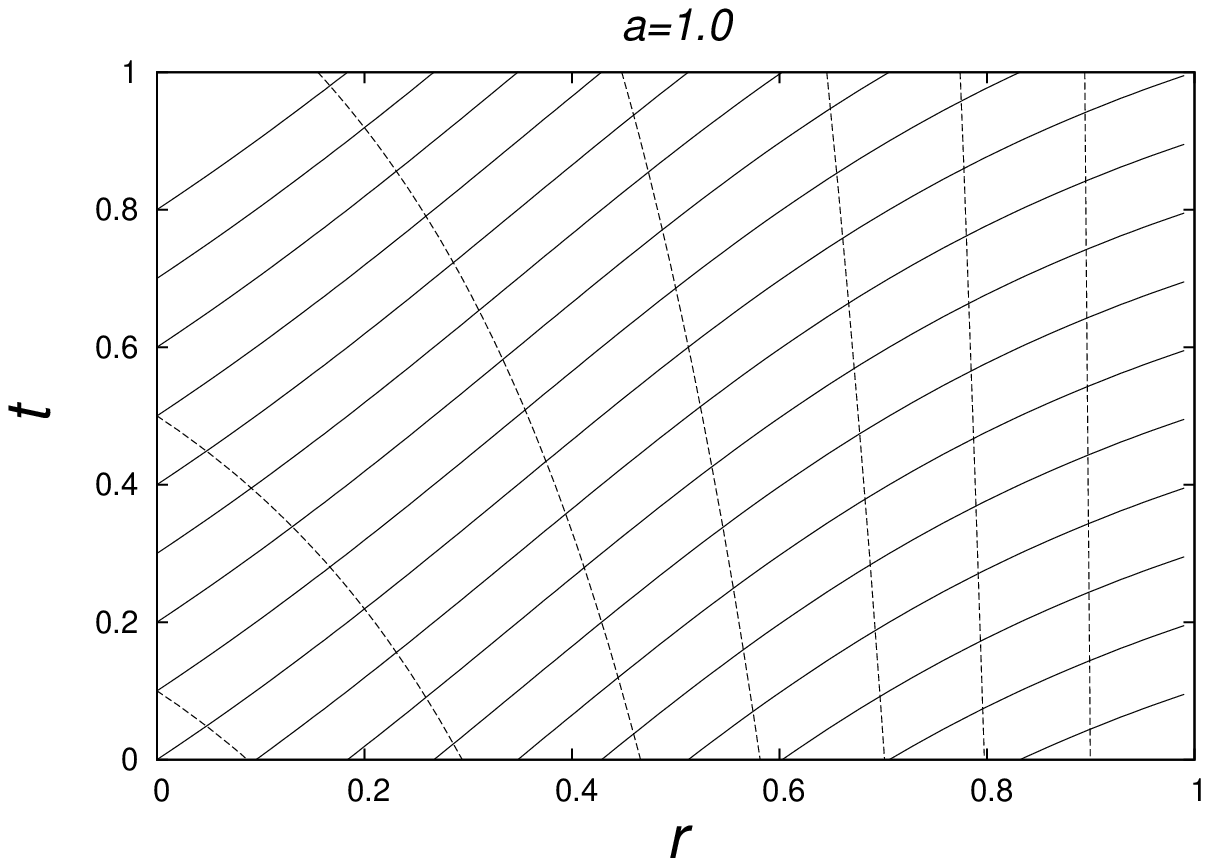}
\includegraphics[width=5cm]{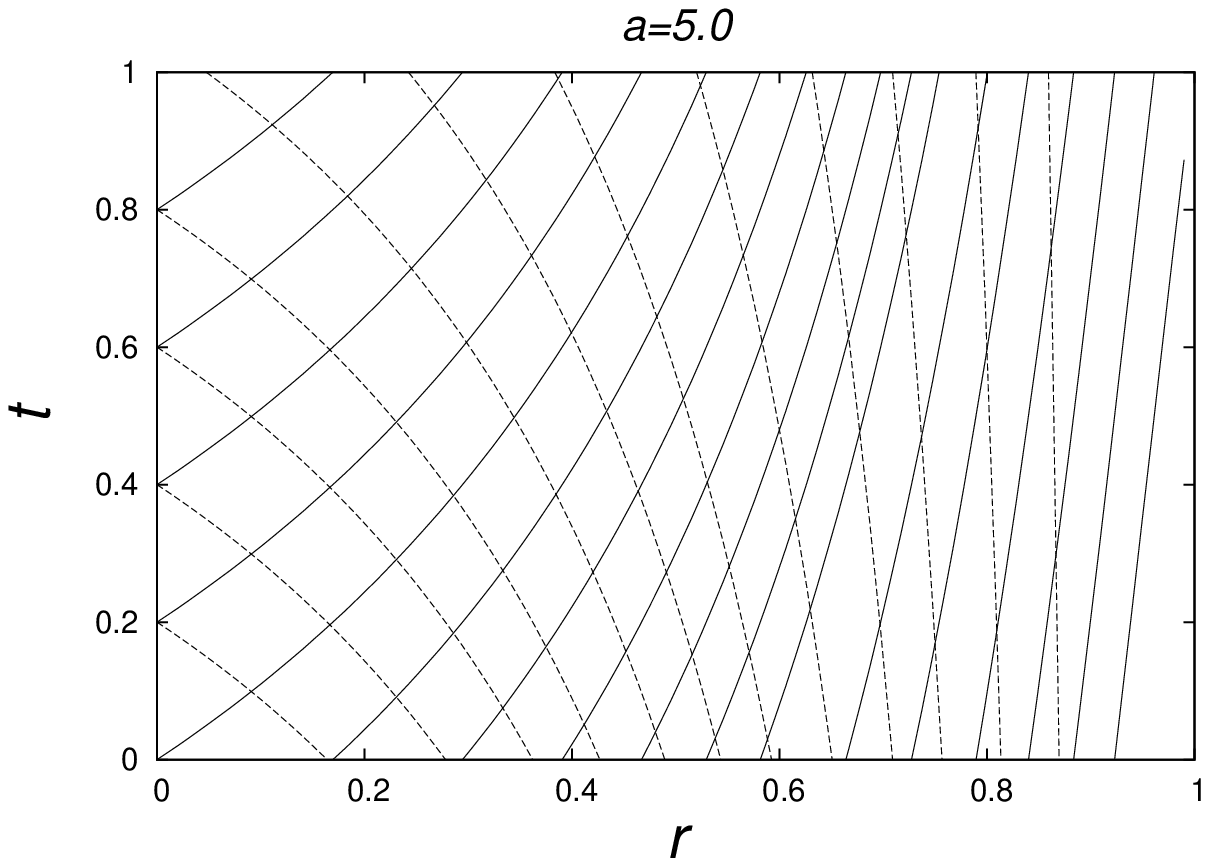}
\caption{\label{fig:brutal_Mink} Space-time diagrams described with hyperboloidal slices and 
$\tilde{k}=6,3,0.6$, or equivalently $a=0.5,1,5$. Continuous (dotted) lines indicate outgoing (in-going) null rays. As in the previous cases, the light cones close near the boundary for small values of the curvature.}
\end{figure*}

\begin{figure*}[ht]
  \centering 
     \flushright 
    \psfrag{ip}{$i^+$} \psfrag{im}{$i^-$} \psfrag{i0}{$i^0$}
     \psfrag{scrp}{$\scri^+$}\psfrag{scrm}{$\scri^-$}
\includegraphics[width=5cm]{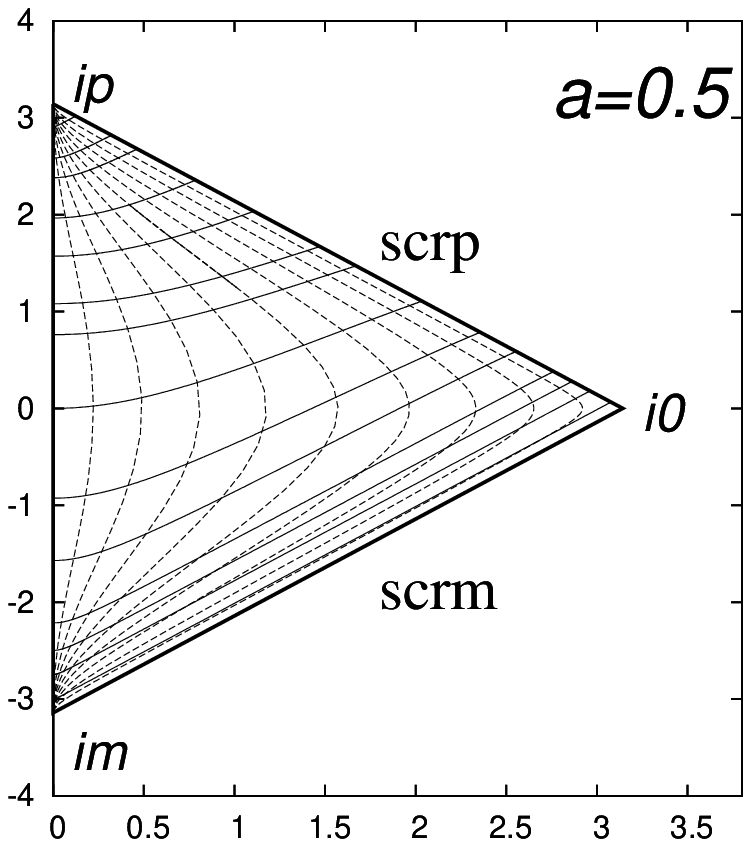}
\includegraphics[width=5cm]{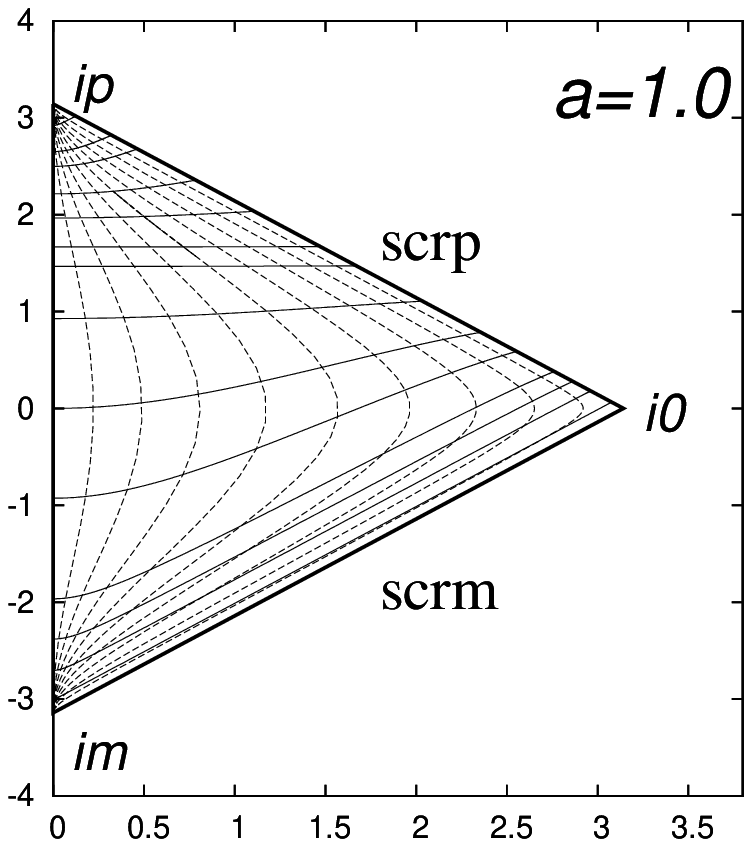}
\includegraphics[width=5cm]{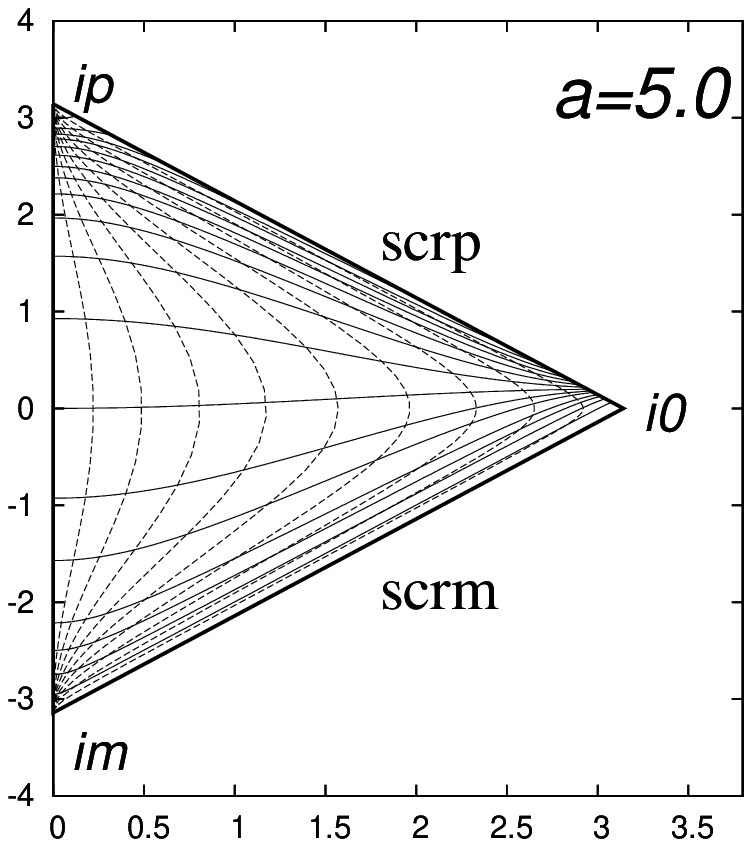}
\caption{\label{fig:conformal_Mink3+1} Conformal diagrams of the spherically symmetric Minkowski space-time in spherical coordinates, using hyperboloidal slices and conformal compactification with $\Omega=1-r$ and various values of $a=0.5,1,5$ or $\tilde{k}=6,3,0.6$. These diagrams are quite similar to those of the 1+1 Minkowski space-time, however notice that we are using a different conformal factor and that the curvature of the slices is different in terms of the parameter $a$. In fact, it has been found that a conformal factor $\Omega=(1-r^2)/2$ is helpful to obtain a regular Ricci scalar at the origin, which is useful to solve the conformally invariant wave equation \cite{anilphi3,bosonstars}.}
\end{figure*}

\noindent The results are presented in Fig. \ref{fig:brutal_Mink} for values $a=0.5,1,5$ or equivalently $\tilde{k}=6,3,0.6$. The corresponding conformal diagrams are shown in Fig. \ref{fig:conformal_Mink3+1}, where the injection of the slices in the space-time depending on the value of the extrinsic curvature is manifest.

\section{Final comments}

We presented the numerical solution of the wave equation on top of particular space-times using hyperbolidal slices and scri-fixing conformal compactification. We calculate the solution as an initial value problem using first order variables, and verified that the resulting constraints when the first order variables are defined converge, and also that the numerical solutions self-converge.

We presented a detailed step-by-step construction of the foliation and conformal compactification for the 1+1 Minkowski space-time. Following previous results we constructed the case of Schwarzschild. We generalized in detail the scri-fixing conformal compactification for the case of a spherically symmetric static space-time, and provide a recipe to construct the adequate coordinates in that case.
We illustrate each case with space-time and conformal diagrams in order to show the global structure of the space-time and how the slices inject into the space-time.

The solution of the wave equation for the case of the Minkowski space-time in 1+1, is in fact  the solution on the physical metric, because the wave operator is conformally invariant. In the Schwarzschild case -four dimensional- we want to stress that we solve the wave equation on top of the conformal space-time background. We studied the case of an outgoing pulse and tracked the late-time behavior of the amplitude of the wave function, and were able to study the quasinormal mode frequencies for various masses of the black hole and modes $l=0,1,2$, and the exponents of the tail decay. We fitted the frequencies found previously and showed that the tail decay exponents obey the restrictions studied in the past.


\section*{Acknowledgments}

We thank Anil Zenginoglu for providing important clues on the scri-fixing conformal compactification. 
This research is partly supported by grants 
CIC-UMSNH-4.9 and 
CONACyT 106466.


\end{document}